\documentclass[prb,twocolumn,superscriptaddress,showpacs,amsmath,amssymb]{revtex4}
\usepackage{float}
\usepackage{amsmath}
\usepackage{graphicx}
\usepackage{hyperref}
\usepackage{bm}
\usepackage[usenames]{color}
\usepackage{verbatim}
\newcommand{\be}{\begin{equation}}
\newcommand{\ee}{\end{equation}}   
\newcommand{\bea}{\begin{eqnarray}}
\newcommand{\eea}{\end{eqnarray}}
\newcommand{\phrl}[1]{Phys.~Rev.~Lett. {\bf #1}}
\newcommand{\phrb}[1]{Phys.~Rev.~B {\bf #1}}
\newcommand{\phrx}[1]{Phys.~Rev.~X {\bf #1}}
\newcommand{\cmat}[1]{arXiv:{\bf #1}}

\newcommand{\RMP}[1]{Rev.~ Mod.~ Phys. {\bf #1}}
\newcommand{\jpcm}[1]{J.~Phys.:Condens.~Matter.{\bf #1}}
\newcommand{\bib}{\bibitem}
\newcommand{\lb}{\left[}
\newcommand{\rb}{\right]}
\newcommand{\lp}{\left(}
\newcommand{\rp}{\right)}

\newcommand{\G}{{\cal G}}
\newcommand{\p}{\mathbf{p}}

\renewcommand{\k}{\mathbf{k}}
\newcommand{\bk}{\bar{k}}

\newcommand{\bGamma}{\bar{\Gamma}}
\newcommand{\bOmega}{\bar{\Omega}}
\newcommand{\bmu}{\bar{\mu}}
\newcommand{\bb}{\bar{b}}
\newcommand{\bepsilon}{\bar{\epsilon}}
\newcommand{\brho}{\bar{\rho}}

\newcommand{\tk}{\widetilde{k}}
\newcommand{\tm}{\widetilde{m}}

\newcommand{\tOmega}{\widetilde{\Omega}}
\newcommand{\tmu}{\widetilde{\mu}}
\newcommand{\tb}{\widetilde{b}}
\newcommand{\tepsilon}{\widetilde{\epsilon}}
\newcommand{\trho}{\widetilde{\rho}}

\newcommand{\Y}{\mathfrak{Y}}
\newcommand{\M}{\mathfrak{M}}
\newcommand{\N}{\mathfrak{N}}
\renewcommand{\G}{\mathfrak{G}}

\begin{document}

\title{Optical response in Weyl semimetal in model with gapped Dirac phase}

\author{S. P. Mukherjee}
\affiliation{Department of Physics and Astronomy, McMaster University, Hamiltion, Ontario, Canada L8S 4M1}

\author{J. P. Carbotte}
\affiliation{Department of Physics and Astronomy, McMaster University, Hamiltion, Ontario, Canada L8S 4M1}
\affiliation{Canadian Institute for Advanced Research, Toronto, Ontario, Canada M5G 1Z8}

\begin{abstract}
We study the optical properties of Weyl semimetal (WSM) in a model which features, in addition to the usual term describing isolated Dirac cones proportional to the 
Fermi velocity $v_{F}$, a gap term $m$ and a Zeeman spin-splitting term $b$ with broken time reversal symmetry. Transport is treated within Kubo formalism and particular 
attention is payed to the modifications that result from a finite $m$ and $b$. We consider how these modifications change when a finite residual scattering rate 
$\Gamma$ is included. For $\Gamma<m$ the A.C. conductivity as a function of photon energy $\Omega$ continues to display the two quasilinear energy regions of 
the clean limit for $\Omega$ below the onset of the second electronic band which is gapped at ($ m+b $). For $\Gamma$ of the order $m$ little trace of two distinct linear 
energy scales remain and the optical response has evolved towards that for $m=b=0$. Although some quantitative differences remain there are no qualitative differences. The 
magnitude of the D.C. conductivity $\sigma^{DC}(T=0)$ at zero temperature ($T=0$) and chemical potential ($\mu=0$) is altered. While it remains proportional to $\Gamma$ 
it becomes inversely dependent on an effective Fermi velocity out of the Weyl nodes equal to $v_{F}^\ast=v_{F}\sqrt{b^2-m^2}/b$ which decreases strongly as the phase boundary 
between Weyl semimetal and gapped Dirac phase (GDSM) is approached at $b=m$. The leading term in the approach to $\sigma^{DC}(T=0)$ for finite $T/\Gamma$, $\mu/\Gamma$ and 
$\Omega/\Gamma$ is found to be quadratic. The coefficient of these corrections tracks closely the $b/m$ dependence of the $\mu=T=\Omega=0$ limit with differences largest 
near to the WSM-GDSM boundary. 

\end{abstract}

\pacs{72.15.Eb, 78.20.-e, 72.10.-d}

\maketitle

\section{Introduction}
\label{sec:I}

By breaking time reversal symmetry a degenerate pair of Dirac mode with linear dispersion curves can be split into a pair of Weyl nodes \cite{Hook,Wan,Balents,Burkov,Hosur}
displaced in momentum space. This displacement determines the Hall conductivity \cite{Burkov} in units of $e^2/\hbar$. The Weyl points have opposite chirality and non
trivial topology and Berry curvature \cite{Xiao} which can play an important role in D.C. transport, in optical properties \cite{Lundgren,Sasaki,Sharma,Nicol,Tabert} as well
as on other properties \cite{Koshino,Burkov1} and their surfaces feature Fermi arcs \cite{Potter}. Examples of experimentally known Weyl semimetals are TaAs and TaP
\cite{Lv,Xu,Weng,Belopolski}. The A.C. optical conductivity in TaAs \cite{Dai} has revealed the expected linear dependence of its interband background in photon energy 
and a $T^2$ dependence of its Drude optical spectral weight. More recently the idea of type II Weyl semimetals with tilted Dirac cones has been introduced \cite{Soluyanov}.
These have distinct optical optical features \cite{Zyuzin,Carbotte}. There is evidence for their existence in TaIrTe$_4$ \cite{Haubold}. Here we will limit our discussion to the 
A.C. optical and D.C. transport properties of a Weyl semimetal without tilt but in a model which contains in addition to a Weyl phase, a gapped Dirac phase.

Our calculations are based on a $4\times 4$ matrix continuum low energy Hamiltonian which has often been utilized in the literature \cite{Burkov,Tabert,Koshino} to model
Weyl semimetals. There are three parameters, the carrier Fermi velocity ( $v_{F}$), a gap($m$) and a Zeeman splitting field($b$). The carrier dispersion curves have two 
branches denoted by $s'=\pm$ and each branch has both a valence ($s=-1$) and conduction($s=1$) band. The carrier energy $\epsilon_{ss'}(\k)$ as a function of momentum $\k$ 
reduce to two identical copies of the simple isotropic Dirac dispersion $\epsilon_{ss'}(\k)=sv_{F}|\k|$ (we have set $\hbar=1$) in the case $m=b=0$. For $b=0$ but $m\ne0$ 
we get instead gapped Dirac cone with $\epsilon_{ss'}(\k)=s\sqrt{v_{F}^2 k^2 + m^2}$. When $b$ and $m$ are both different from zero there are two distinct phases. If $b>m$ 
we get a Weyl semimetal while for $b<m$ a gapped Dirac phase emerges. In the Weyl phase the nodes are at momentum $k_{z}=\pm\sqrt{b^2-m^2}/v_{F}$ for $k_{x}=k_{y}=0$ for 
the $s'=-1$ branch while the $s'=1$ branch has a gap of magnitude ($m+b$). In the gapped Dirac phase the $s'=-1$ branch has a gap of ($m-b$) while for the $s'=1$ 
remains at ($m+b$).

In this paper we start with an expression for the absorptive part of the longitudinal dynamic optical conductivity $\sigma_{xx}(T,\Omega)$, a function of temperature $T$
and photon energy $\omega$ based on our model Hamiltonian and an associated Kubo formula. From previous works \cite{Lundgren,Sasaki,Sharma,Nicol} it is known that details 
of the model used to treat disorder can significantly affect the optical response. For example in references (\onlinecite{Lundgren} and \onlinecite{Nicol}) in the case of 
the simplest Dirac cone with $m=b=0$ and electron dispersion $\epsilon_{ss'}(\k)=sv_{F}|\k|$, three models of residual impurity scattering rates were employed. The simplest 
was a constant rate $\Gamma$, the second weak scattering in Born approximation for which $\Gamma$ is proportional to energy squared ($\omega^2$) and charged impurities 
with $\Gamma$ inversely proportional to $\omega^2$. The aim of the present study is to provide a first understanding of any essential difference introduced in optics and 
D.C. transport when a finite gap $m$ and Zeeman term $b$ are introduced in the Hamiltonian. For this purpose it is sufficient to use the simplest constant $\Gamma$ model.
This is consistent with work of Holder et. al.\cite{Holder} who argue that the density of state at the Weyl point is  always finite.
In section II we present the necessary formalism and give results for the optical conductivity at $T=0$ as a function of the photon energy $\Omega$ for various well chosen 
values of $b/m$, $\Gamma$ and chemical potential $\mu$. In section III we derive simple analytic algebraic expressions for the D.C. conductivity $\sigma^{D.C.}$ at 
$T=\mu=0$ (charge neutrality) and find that, the known formula \cite{Nicol} for the $m=b=0$ case which finds $\sigma^{D.C.}$ to be directly proportional to $\Gamma$ and 
inversely proportional to the Fermi velocity $v_{F}$ still holds but $v_{F}$ to be replaced by the effective Fermi velocity $v_{F}^\ast$ at the Weyl nodes which is, for 
$\bb=b/m>1$, $v_{F}\sqrt{b^2-m^2}/b$. The linear in $\Gamma$ and inversely proportional to $v_{F}^\ast=v_{F}\sqrt{b^2-m^2}/b$ is to be contrasted to the 2-D case for 
which, in the same constant $\Gamma$ approximation, $\sigma^{D.C.}_{2D}$ is found to be $\frac{4e^2}{\pi h}$ \cite{Sharapov,Tan,Katsnelson}, universal independent of any 
material parameters and in particular scattering rate $\Gamma$. A universal minimum conductivity does arise in other contexts. For example in d-wave superconductors \cite{Lee} 
this remains true even when the gap symmetry \cite{Donovan,Donovan1,Hwang} goes beyond the simplest $d_{x^2-y^2}$ \cite{Jiang} model provided the gap goes through 
zero \cite{Schachinger}. It does not occur however for isotropic s-wave gap symmetry even when there is anisotropy but no zero \cite{Leung}. It does however arise in the 
underdoped regime of the cuprate \cite{Carbotte1} superconductors where a pseudogap emerges which effectively provides an important energy dependence to the underlying 
normal state density of state \cite{Mitrovic}. Other related works can be found in the literature \cite{Roy, Ominato, Mirlin}.
In section-IV we consider leading order corrections to $\sigma^{D.C.}_{min}$ due to finite temperature and chemical potential. Both corrections are found to be quadratic 
in $T/\Gamma$ and $\mu/\Gamma$. Section V deals with the finite $\Omega$ approach which is also of order $(\Omega/\Gamma)^2$. A summary and conclusions appear in 
section-VI.

\section{Formalism and $T=0$ optical conductivity}
\label{sec:II}

We consider the $4\times 4$ matrix continuum Hamiltonian of the form
\be
\label{Basic Hamiltonian}
\hat{H}=v_{F} \hat{\tau}_x \lp \hat{\mathbf{\sigma}}.\p\rp +m \hat{\tau}_z + b \hat{\sigma}_z,
\ee
used before by Koshino and Hizbullah \cite{Koshino} to discuss magnetization and Tabert and Carbotte \cite{Tabert} who considered the A.C. optical conductivity in clean 
limit. In Eq.~(\ref{Basic Hamiltonian}) $v_{F}$ is the Fermi velocity, $m$ is a mass and $b$ describes an intrinsic Zeeman field characteristic of a magnetic which breaks 
time reversal symmetry. Besides these three material dependent parameters ($\hat{\tau}_x,\hat{\tau}_y,\hat{\tau}_z$) are a set of $2\times 2$ Pauli matrices related to 
pseudospin while ($\hat{\sigma}_x,\hat{\sigma}_y,\hat{\sigma}_z$) are a second set related to electron spin. Finally $\k$ is momentum. There are 4 bands of which two are 
conduction bands and two are valence bands with the two branches denoted by $s'=\pm$. The dispersion curves
\be
\label{Dispersion}
\epsilon_{ss'}(\k)=s\sqrt{\lp k^2_{x}+k^2_{y}\rp + \lp \sqrt{k^2_{z}+m^2} + s'b\rp^2}
\ee
with $s=\pm$, the plus gives the conduction and the minus the valence band associated with the branch $s'=\pm$. If we normalize momentum $\k$ by $m$, in terms of 
$\bar{k}=\k/m$ we have
\be
\label{Dispersion-normalized}
\frac{\epsilon_{ss'}(\k)}{m}=s\sqrt{\lp \bk^2_{x}+\bk^2_{y}\rp + \lp \sqrt{\bk^2_{z}+1} + s'\bb\rp^2}=s\bepsilon_{s'}(\k).
\ee

In the appendix we provide expressions for the interband background $\sigma^{IB}(T=0,\Omega)$ and $\sigma^{D}(T=0,\Omega)$ valid for finite photon energy $\Omega$ at zero 
temperature and constant residual scattering $\Gamma$. The expression for the interband conductivity normalized to $\Gamma$ is
\be
\label{EqA5-appendix}
\frac{\sigma^{\hspace{-0.1cm}IB}_{\hspace{-0.1cm}xx}\hspace{-0.1cm} \lp\hspace{-0.1cm}T\hspace{-0.1cm}=\hspace{-0.1cm}0,\hspace{-0.07cm}\Omega\rp}{\Gamma}\hspace{-0.1cm}=\hspace{-0.1cm} 
\frac{e^2}{2\pi^3 \hbar^2v_{F}}\hspace{-0.15cm}\sum_{s'} \hspace{-0.2cm}\int^{\hspace{-0.05cm}\infty}_{\hspace{-0.05cm}0}\hspace{-0.2cm}
\frac{d\tk_{z}}{\tOmega}\hspace{-0.2cm}\int^{\hspace{-0.05cm}\infty}_{\hspace{-0.05cm}0}\hspace{-0.35cm} \trho d\trho 
\hspace{-0.1cm}\lp\hspace{-0.15cm}1\hspace{-0.1cm}-\hspace{-0.1cm}\frac{\brho^2}{2\tepsilon^2_{s'}}\hspace{-0.15cm}\rp \hspace{-0.13cm}\Y(\tmu,\tOmega,\tb,\tm,\tepsilon_{s'}), 
\ee
This is Eq.(\ref{IntIB}) with the explicit algebraic function $\Y(\tmu,\tOmega,\tb,\tm,\tepsilon_{s'})$ given by Eq.(\ref{Special-function-interband}) and not repeated 
here. Similarly for the Drude contribution conductivity equation (\ref{OC Drude}) applies which is 
\be
\label{EqA6-appendix}
\frac{\sigma^{D}_{xx}(T\hspace{-0.1cm}=\hspace{-0.1cm}0,\Omega)}{\Gamma}=\frac{e^2}{2\pi^3 \hbar^2v_{F}}\hspace{-0.15cm} \sum_{s'}\hspace{-0.15cm} 
\int^{\infty}_{0} \hspace{-0.15cm}\frac{d\tk_{z}}{\tOmega} \hspace{-0.2cm}\int^{\infty}_{0}\hspace{-0.4cm} \trho d\trho 
\frac{\trho^2}{2\tepsilon^2_{s'}} \mathfrak{H}(\tmu,\tOmega,\tb,\tm,\tepsilon_{s'}), 
\ee
\begin{figure}
\includegraphics[width=2.5in,height=3.2in, angle=270]{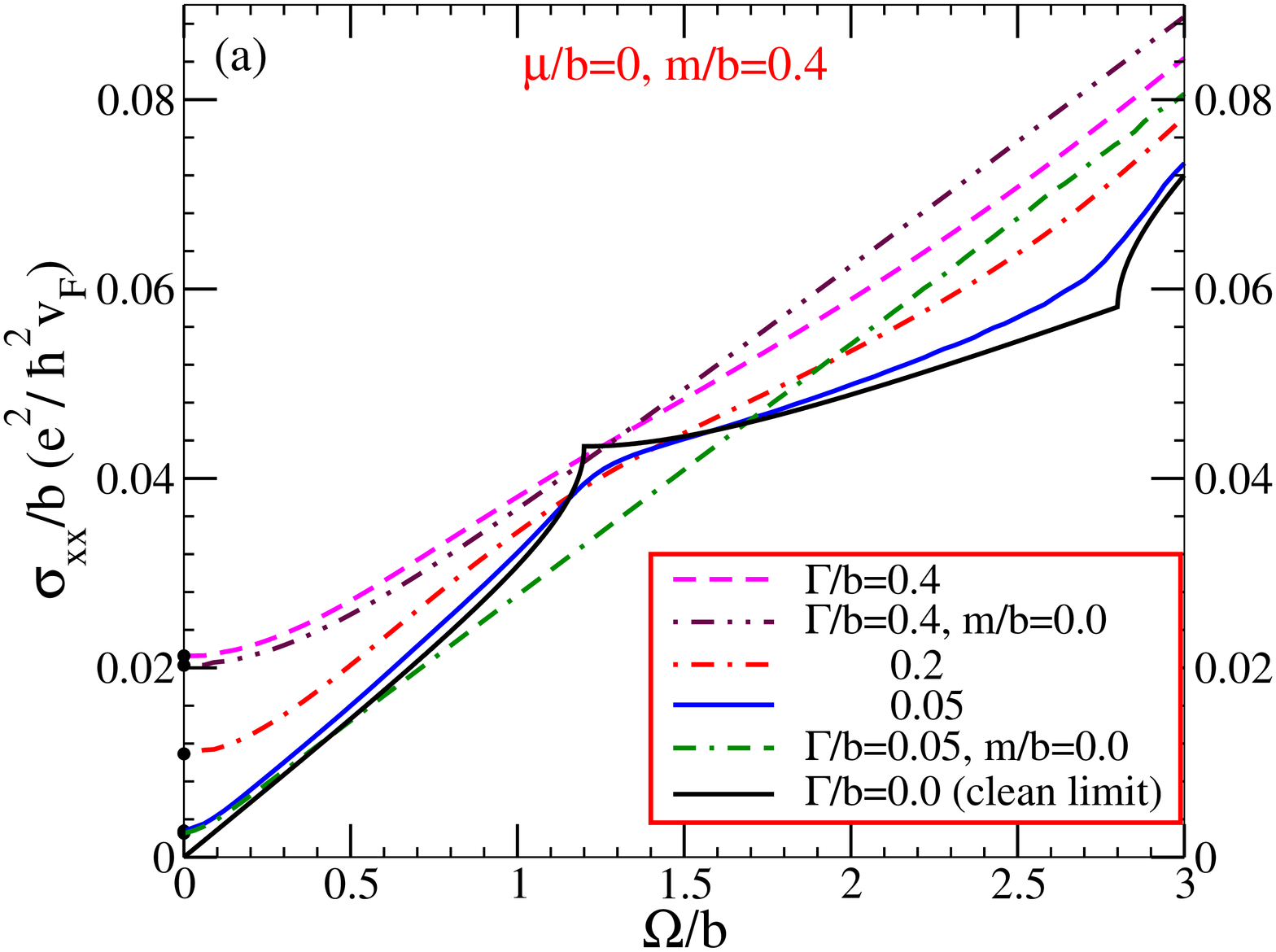} 
\includegraphics[width=2.5in,height=3.2in, angle=270]{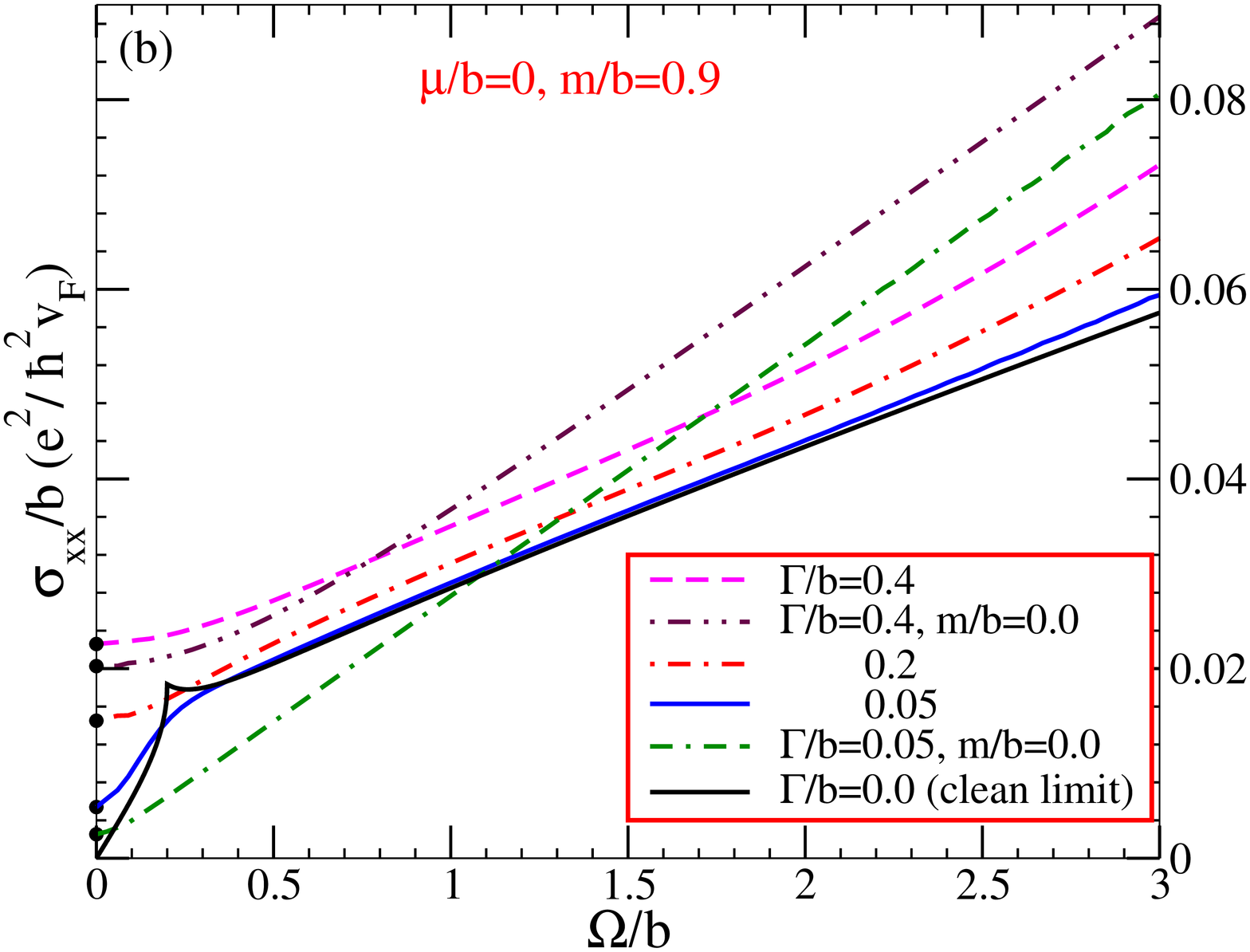} 
\caption{(Color online) The A.C. optical conductivity $\sigma_{xx}(T=0,\Omega)$ at $T=0$ normalized to $b$ in units of $\frac{e^2}{\hbar^2v_{F}}$ as a function of 
photon energy $\Omega$ also normalize to $b$ for four values of the residual scattering $\Gamma/b$. The solid black curve is the clean limit result of 
Ref.[\onlinecite{Tabert}] and is included here for comparison with the finite $\Gamma/b$ results namely solid blue curve for $\Gamma/b=0.05$, dashed-dotted red curve for 
$\Gamma/b=0.2$ and dashed purple curve for $\Gamma/b=0.4$. In all curves the solid black dots on the $\Omega/b=0$ axis are the D.C. values obtained from 
Eq.(\ref{DC-cond-final}). Frame (a) is for $m/b=0.4$ away from phase boundary and (b) for $m/b=0.9$ is closer to the GDSM-WSM boundary at $b/m=1.0$. In both cases the 
chemical potential is zero $\mu=0$ (charge neutrality).}
\label{fig:Fig1}
\end{figure}
with $\mathfrak{H}(\tmu,\tOmega,\tb,\tm,\tepsilon_{s'})$ another explicit algebraic function given in Eq.(\ref{Special-function-intraband}). The remaining double 
integral over $\tk_{z}$ and $\trho $ needs to be done numerically. In Fig.~\ref{fig:Fig1} we present results for $\sigma_{xx}(T=0,\Omega)$ at $T=0$ normalized to $b$ in 
units of $\frac{e^2}{\hbar^2v_{F}}$ as a function of photon energy $\Omega$ also normalize to $b$. Both frames are for zero value of the chemical potential $\mu/b=0$. 
The top frame is for $m/b=0.4$ and the bottom frame for $m/b=0.9$ closer to the boundary of the gapped Dirac phase at $m/b=1$. 

\begin{figure}
\includegraphics[width=2.5in,height=2.5in, angle=270]{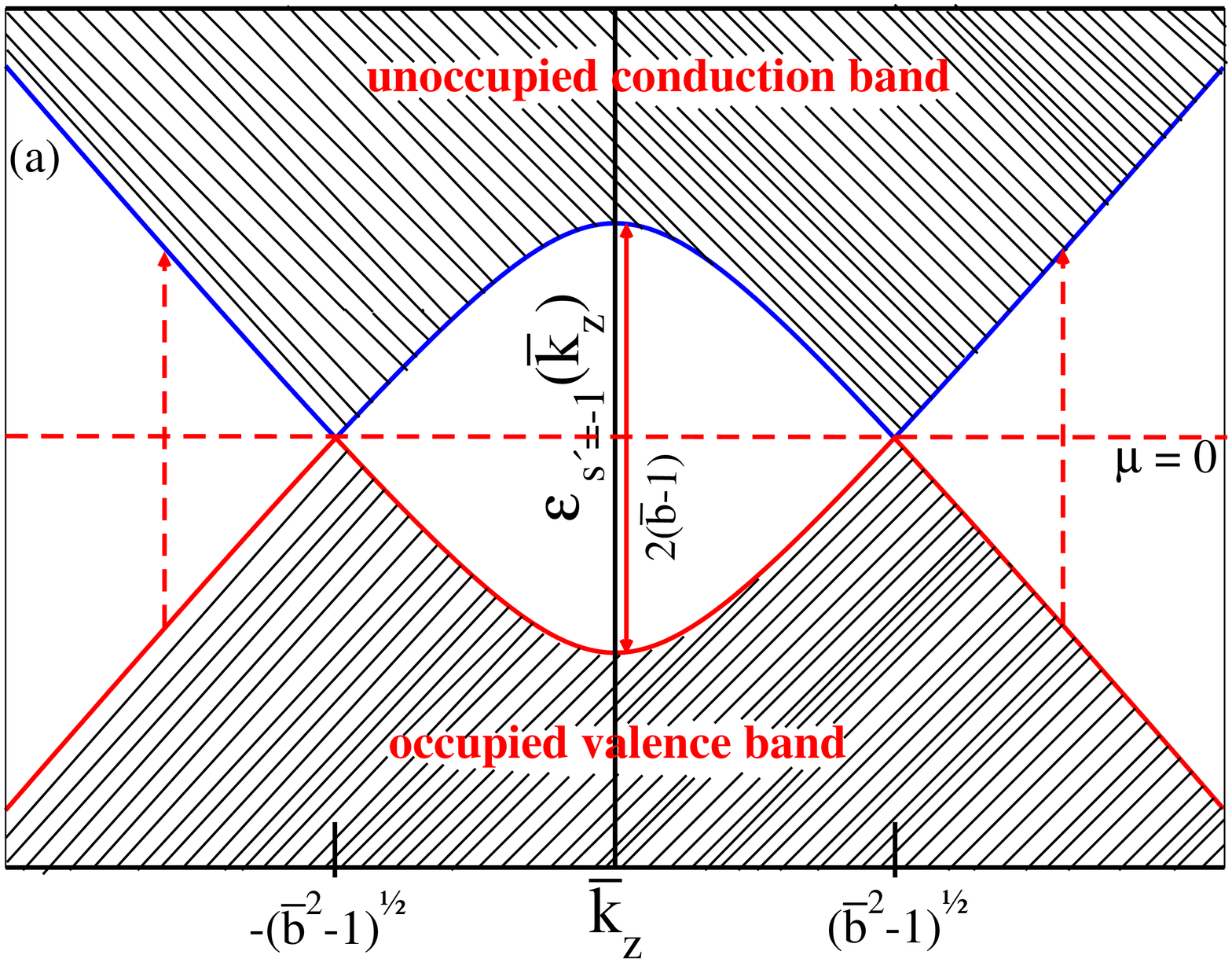} 
\includegraphics[width=2.5in,height=2.5in, angle=270]{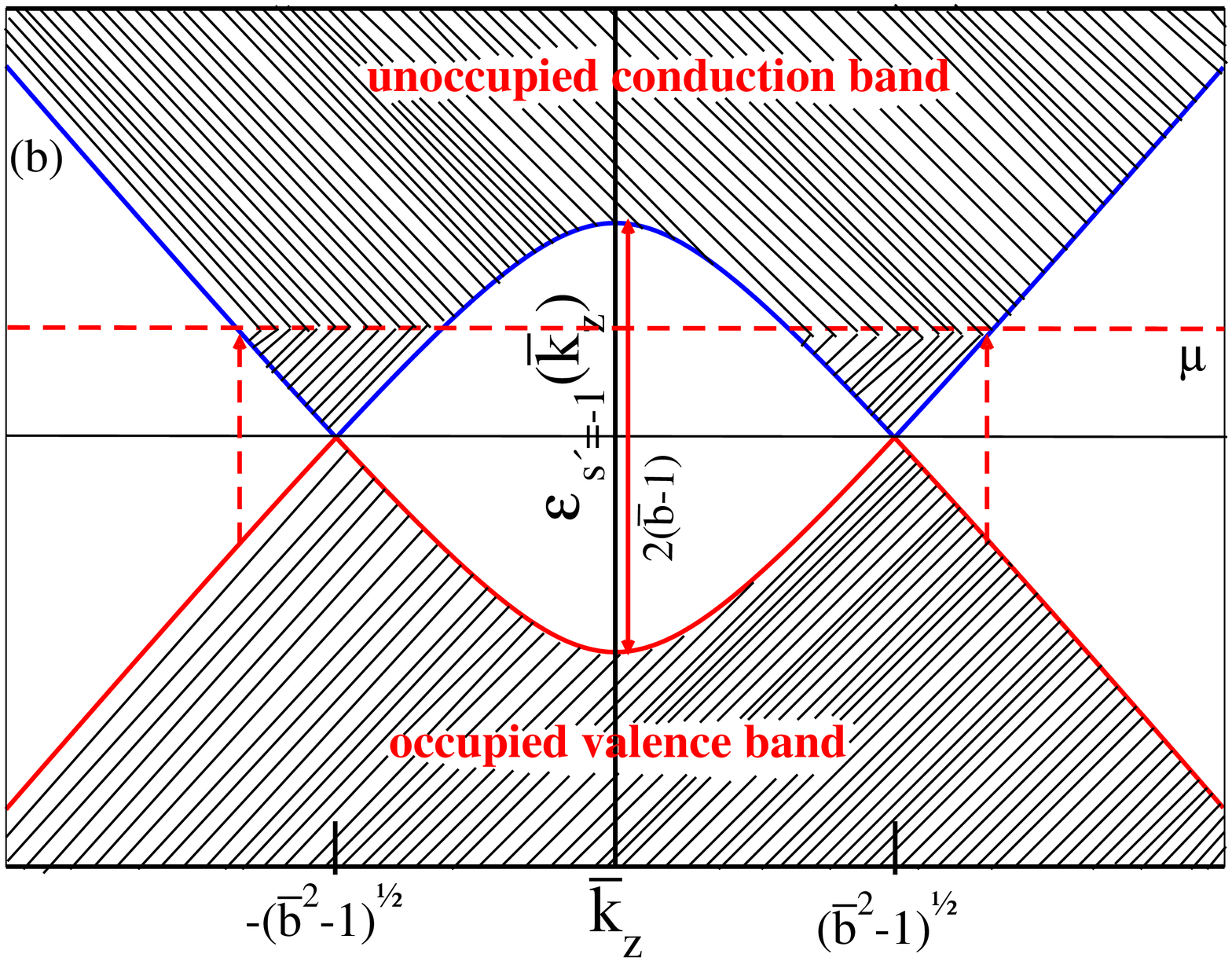} 
\caption{(Color online) The electronic dispersion curves for $k_{x}=k_{y}=0$ branch $\bepsilon_{s'=-1}(\bk_{z})$ normalized to $m (\bk_{z}=k_{z}/m)$ as a function of 
$z$-component of momentum $k_{z}$ also normalized to $m (\bk_{z}=k_{z}/m)$ with $k_{x}=k_{y}=0$. The Weyl nodes are at $\bk_{z}=\pm\sqrt{\bb^2-1}$. The dashed red arrows 
show possible interband optical transitions. The energy of such a transition for $\bk_{z}=0$ which involves a Van Hove singularity has magnitude $2\lp\bb-1\rp$ and is 
shown as a solid red arrow. The top frame is for $\mu=0$ (charge neutrality) and the bottom frame is a dopped case with finite chemical potential $\mu$. The shaded region 
below the chemical potential level $\mu$ in the conduction band shows the Pauli blocked region where no interband transitions are possible for $\bOmega<2\bmu$. The lost 
optical spectral weight in the interband background is transferred to the intraband transitions which manifest as a Drude (see Fig.~\ref{fig:Fig3}).}
\label{fig:Fig2}
\end{figure}

In both frames the solid black curve is the clean limit result of Ref.[\onlinecite{Tabert}](see their Fig3 upper frame). In this case the D.C. limit of the conductivity 
is zero. The solid blue curves are for finite $\Gamma/b=0.05$, dashed-dotted red curve for $\Gamma/b=0.3$ and dashed purple curve for $\Gamma/b=0.4$. The two additional 
curves double dashed-dotted (green) has $\Gamma/b=0.05$ and double dotted-dashed (black) has $\Gamma/b=0.4$ and are for comparison. Here $m/b=0$ (simple Dirac nodes) and 
the optical response is basically linear in photon energy $\Omega$ except for a turn up as $\Omega$ gets small towards a finite value of D.C. conductivity shown as solid 
black dots. This limit will be discussed in detail in the next section. Comparing black and blue curves we note that the introduction of a small residual scattering 
$\Gamma/b=0.05$ broadens slightly the Van Hove singularity of the clean limit at $\Omega/b=1.2$ as well as the second singularity at $\Omega/b=2.8$ (black curve). The first 
singularity corresponds to the energy of the maximum interband optical transition possible along $k_{z}=0$ as illustrated schematically in Fig.~\ref{fig:Fig2}. Here we 
show the electronic dispersion curves for the ungapped branch for $k_{x}=k_{y}=0$ as a function of $\bk_{z}$. Optical transitions are indicated by vertical red arrows. 
These are possible on either sides of the two Weyl nodes at $\bk_{z}=\pm\sqrt{\bb^2-1}$. The magnitude of such transitions is not limited for 
$|\bk_{z}|>\pm\sqrt{\bb^2-1}$ and can extend to large photon energies. However those for $|\bk_{z}|<\pm\sqrt{\bb^2-1}$ cannot be larger than 
$\frac{\Omega}{b}=2\lp 1-m/b \rp$ which occurs at $\bk_{z}=0$ (Fig.~\ref{fig:Fig2} solid red arrow). The second singularity at $\Omega/b=2.8$ is due to the onset of the 
optical transition coming from the second branch of the dispersion curves which provides an optical gap of $\frac{\Omega}{b}=2\lp 1+m/b \rp$. Returning to the top frame 
of Fig.~\ref{fig:Fig1} the blue curve with $\Gamma/b=0.05$ displays a broaden shoulder at $\Omega/b=1.2$ and a broaden elbow at $\Omega/b=2.8$. The two distinct quasilinear 
regions for photon energies between 0 and $2(b-m)$ and between $2(b-m)$ to $2(b+m)$ which has the smaller slope, remain very well defined. Such features are not present 
in double dashed-dotted curve for which $m/b=0$ and the optical response is basically linear in $\Omega$ with a small turn up at the $\Omega\to 0$ limit.

As the broadening is increased to $\Gamma/b=0.2$ (dashed dotted red) a small broad shoulder does remain around $\Omega=2(b-m)$ and a very broaden elbow around $\Omega=2(b+m)$ 
but the two quasilinear regions of the clean limit with distinct slopes are no longer prominent. When $\Gamma/b=0.4$ the dashed purple applies and now the optical response 
has evolved to be much closer to that of $m/b=0$ (simple Dirac) shown as the double dotted-dashed brown curve where a single slope is clearly manifest. Careful comparison 
of these two sets of results show some remaining quantitative but no qualitative differences. Similar remarks apply to the lower frame of Fig.~\ref{fig:Fig1} where $m/b=0.9$ 
much closer to the WSM-GDSM phase boundary at $m/b=1$. An important difference to note and we will elaborate on this in the next two sections, is the D.C. value of the 
conductivity $\sigma^{DC}$ shown as heavy black dots on the vertical axis. When compared with the $m/b=0$ case there is a significant increase in the magnitude of $\sigma^{DC}$
due to finite $m$. 
 
\begin{figure}
\includegraphics[width=2.5in,height=3.2in,angle=270]{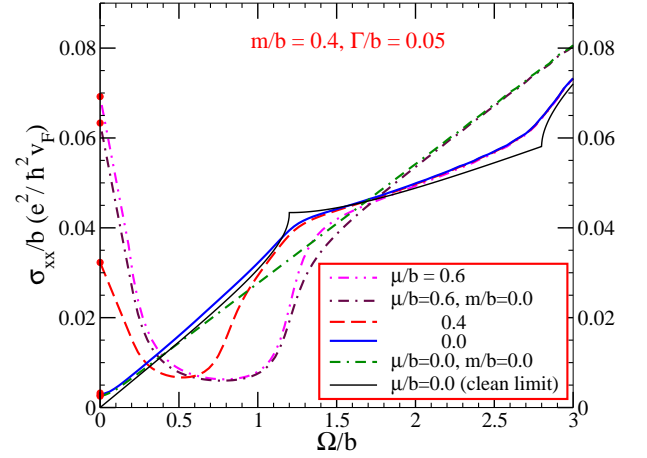} 
\caption{(Color online) The A.C. optical conductivity $\sigma_{xx}(T=0,\Omega)$ at $T=0$ normalized to $b$ in units of $\frac{e^2}{\hbar^2v_{F}}$ as a function of 
photon energy $\Omega$ also normalize to $b$ for four values of the chemical potential $\mu/b$. The solid black curve is for comparison is the clean limit result of 
Ref.[\onlinecite{Tabert}] with $\mu/b=0$. The solid blue curve is for $\mu/b=0.0$ with small residual scattering $\Gamma/b=0.05$. Apart from that the red dashed curve is 
for  $\mu/b=0.4$ and purple dashed-double dotted curve for $\mu/b=0.6$. In all cases $m/b=0.4$ and $\Gamma/b=0.05$ except for the clean limit case. Note the increasing 
prominence of the Drude peak with increasing value of the chemical potential $\mu$.}
\label{fig:Fig3}
\end{figure}

Results in Fig.~\ref{fig:Fig3} are for finite chemical potential. In all cases the scattering rate $\Gamma/m$ is set at 0.05 except for the black curve which 
is the clean limit. The solid blue and double dashed-dotted green are repeated from Fig.~\ref{fig:Fig1} for easy comparison. The dashed red has a chemical potential 
$\mu/b=0.4$ while the purple dashed-double dotted has $\mu/b=0.6$. The additional brown double dotted-dashed is for $m/b=0.0$ (simple Dirac) with $\mu/b=0.6$ for comparison 
with the double dotted-dashed purple curve. These differ only by the value of $m/b$ used. We see little differences in the Pauli block region where the conductivity is small. 
The missing optical spectral weight in the interband background in this region has been transferred to the Drude at small $\Omega$ because of Pauli blocking as illustrated 
in the bottom frame of Fig.~\ref{fig:Fig2}. No optical transition with $\Omega<2\mu$ are now possible. In the region above $\Omega \cong 2\mu$ the background rapidly 
recovers its clean limit value except for a small amount of broadening but this background is very different for finite $m/b$ as compared with the $m/b=0$ case as we have 
already discussed. We make one final point. At $\Omega=0$ the D.C. conductivity is significantly greater for the finite $m/b$ case which will be taken up next.
A recent study \cite{Chinotti} of the longitudinal optical response of YbMnBi$_{2}$ has shown the two energy scales defined by distinct quasilinear absorption regions as 
in our theoretical curve and the Drude peak indicating a rather clean sample (narrow Drude) with finite dopping away from charge neutrality.

\section{D.C. limit at zero temperature ($T=0$) and chemical potential ($\mu=0$)}
\label{sec:III}

The expressions for the dynamic optical conductivity at any temperature $T$ and photon energy $\Omega$ are given in the Appendix. They are a generalization to include self energy effects in the work of 
reference \onlinecite{Tabert} which was valid only in clean limit. The D.C. limit takes the form 
\bea
\label{DCD}
&& \sigma^{D}_{xx}(T,\Omega=0)= \frac{e^2\pi}{\hbar^2 v_{F}}\sum_{ss'=\pm} \int^{+\infty}_{-\infty}\hspace{-0.5cm} d\omega \lp -\frac{\partial f(\omega)}{\partial \omega}\rp   \int \frac{d^3\k}{(2\pi)^3} \nonumber\\ 
&& \hspace{-0.2cm}\frac{k^2-k^2_z}{2\epsilon^2_{ss'}(\k)}A^2(\epsilon_{ss'},\omega) ,
\eea
for the intraband or Drude contribution (\ref{OCD}). The interband contribution is
\bea
\label{DCIB}
&& \sigma^{IB}_{xx}(\Omega=0)= \frac{e^2\pi}{\hbar^2 v_{F}}\sum_{ss'=\pm} \int^{+\infty}_{-\infty}\hspace{-0.5cm} d\omega \lp -\frac{\partial f(\omega)}{\partial \omega}\rp   \int \frac{d^3\k}{(2\pi)^3} \nonumber\\ 
&& \hspace{-0.2cm}\lb 1-\frac{k^2-k^2_z}{2\epsilon^2_{ss'}(\k)}\rb A(\epsilon_{ss'},\omega) A(-\epsilon_{ss'},\omega),
\eea
with the carrier spectral density $A(\epsilon_{ss'},\omega)$ defined in Eq.~(\ref{spectral-density}) for the case of a constant scattering rate $\Gamma$. At charge neutrality the chemical potential 
$\mu$ is zero and Eq.~(\ref{DCD}) and (\ref{DCIB}) take on a particularly simple form. The sum of $\sigma^D$ plus $\sigma^{IB}$ denoted by $\sigma^{DC}$ add, at zero temperature, such that the factor 
$\frac{k^2-k^2_{z}}{2\epsilon^2_{ss'}}$ drops out and we get,
\be
\label{DC-conductivity}
\sigma^{DC}(\Omega=0)= \frac{2e^2\pi}{\hbar^2 v_{F}}\sum_{s'=\pm} \int \frac{d^3\k}{(2\pi)^3} \lp\frac{\Gamma}{\pi}\rp^2 \lp\frac{1}{\Gamma^2+\epsilon^2_{s'}}\rp^2.
\ee
It is convenient to introduce polar coordinates to treat $k_{x},k_{y}$ degrees of freedom and we get 
\bea
\label{DC-cond-new}
&& \sigma^{DC}(\Omega=0)= \frac{2e^2\pi}{\hbar^2 v_{F}}\sum_{s'=\pm} \int \frac{\rho d\rho}{(2\pi)^2} \int^{+\infty}_{-\infty} dk_z \lp\frac{\Gamma}{\pi}\rp^2 \times \nonumber\\
&& \lp \frac{1}{\Gamma^2+\rho^2+ \lp\sqrt{k^2_z+m^2} + s'b\rp^2}\rp^2.
\eea
The integration over $\rho$ is elementary and yields,
\bea
\label{DC-cond-rho}
&& \sigma^{DC}(\Omega=0)= \frac{e^2\pi}{\hbar^2 v_{F}}\lp\frac{\bGamma}{\pi}\rp^2 \sum_{s'=\pm} \frac{m}{(2\pi)^2} \int^{+\infty}_{-\infty} d\bk_z  \times \nonumber\\
&& \frac{1}{\bGamma^2+\lp\sqrt{\bk^2_z+1} + s'\bb\rp^2}.
\eea
where we have scaled out a factor of $m$ and $\bGamma=\Gamma/m$ $\bb=b/m$. Doing the sum over $s'$ we get,
\bea
\label{DC-cond-summed}
&& \sigma^{DC}(\Omega=0)= \frac{e^2 m\bGamma^2}{2\pi^3 \hbar^2 v_{F}} \int^{+\infty}_{0} d\bk_z  \times \nonumber\\
&& \lb\frac{1}{\bGamma^2+\lp\sqrt{\bk^2_z+1} + \bb \rp^2}+\frac{1}{\bGamma^2+\lp\sqrt{\bk^2_z+1} - \bb \rp^2}\rb.
\eea
There is no analytic solution to the final integral over $\bk_z$ in Eq.~(\ref{DC-cond-summed}) and we need to proceed numerically. This integral is a function of 
$\bGamma,\bb$ which we denote as $\N(\bGamma,\bb)$ with
\be
\label{mathfracN}
\N(\bGamma,\bb)\hspace{-0.1cm}=\hspace{-0.2cm}\int^{+\infty}_{0}\hspace{-0.5cm}dx \hspace{-0.1cm}\lb\frac{1}{\bGamma^2\hspace{-0.1cm}+\hspace{-0.1cm}\lp\hspace{-0.1cm}\sqrt{x^2\hspace{-0.1cm}+\hspace{-0.1cm}1}\hspace{-0.1cm}+\hspace{-0.1cm}\bb \rp^2}+
\frac{1}{\bGamma^2\hspace{-0.1cm}+\hspace{-0.1cm}\lp\hspace{-0.1cm}\sqrt{x^2\hspace{-0.1cm}+\hspace{-0.1cm}1}\hspace{-0.1cm}-\hspace{-0.1cm}\bb \rp^2}\rb
\ee
and 
\be
\label{DC-cond-final}
\sigma^{DC}(\Omega=0)= \frac{e^2 m\bGamma^2}{2\pi^3\hbar^2 v_{F}} \N(\bGamma,\bb).
\ee
For $b=0$ we simply get two versions of decoupled gapped Dirac cones. In this limit
\be
\label{mathfracY-b0}
\N(\bGamma,\bb)= 2 \int^{+\infty}_{0} \frac{dx}{\bGamma^2+x^2+1}=\frac{\pi}{\sqrt{\bGamma^2+1}}
\ee
and hence
\be
\label{DC-cond-b0}
\sigma^{DC}(\Omega=0)= \frac{e^2}{2\pi^2\hbar^2 v_{F}} \frac{\Gamma^2}{\sqrt{\Gamma^2+m^2}}.
\ee
which for $m=0$ gives the known result of Ref.[\onlinecite{Nicol}] for Dirac cones namely $\sigma^{DC}(\Omega=0)= \frac{e^2 \Gamma}{2\pi^2\hbar^2 v_{F}}$ which is 
proportional to the scattering rate $\Gamma$. When $m\ne0$ and in fact $m>>\Gamma$ we get the $m=0$ value multiplied by a further factor of $\Gamma/m$. This reduces the
value of the minimum conductivity below that of the ungapped case as expected.

A second feature of the dispersion curves given in Eq.~(\ref{Dispersion-normalized}) is that for $m=0$, the material parameter $b$ drops out of the expression for 
$\sigma^{DC}(\Omega=0)$. We note that in that case,
\be
\label{dispersion-m0}
\epsilon_{ss'}=s\sqrt{k^2_x+k^2_y+\lp|k_z|+bs'\rp^2}
\ee
and returning to Eq.~(\ref{DC-cond-new}) we obtain
\bea
\label{DC-cond-m0}
&& \sigma^{DC}(\Omega=0)= \frac{e^2\Gamma^2}{2\pi^3\hbar^2 v_{F}} \int^{+\infty}_{0} dk_z \times \nonumber\\
&& \lp\frac{1}{\Gamma^2+(k_z+b)^2}+\frac{1}{\Gamma^2+(k_z-b)^2}\rp
\eea
In each of the two integrals over $k_z$ we can make a change of variable to $k'_z=k_z+bs'$ to get
\be
\label{DC-cond-change-variable}
\sigma^{DC}(\Omega=0)= \frac{e^2\Gamma^2}{2\pi^3\hbar^2 v_{F}}\biggl\{\int^{+\infty}_{b}\hspace{-0.3cm} \frac{dk'_z}{\Gamma^2+k'^2_z} + \int^{+\infty}_{-b}\hspace{-0.3cm} \frac{dk'_z}{\Gamma^2+k'^2_z}\biggr\}
\ee
The contribution of the lower limits to the integral will cancel and we pick up only the upper limit part which give $\frac{\pi}{2\Gamma}$ twice and we get back 
Eq.~(\ref{DC-cond-b0}) with $m=0$, the known answer \cite{Xiao} for two isolated Dirac nodes. 

\begin{figure}[h]
\includegraphics[width=2.5in,height=3.2in, angle=270]{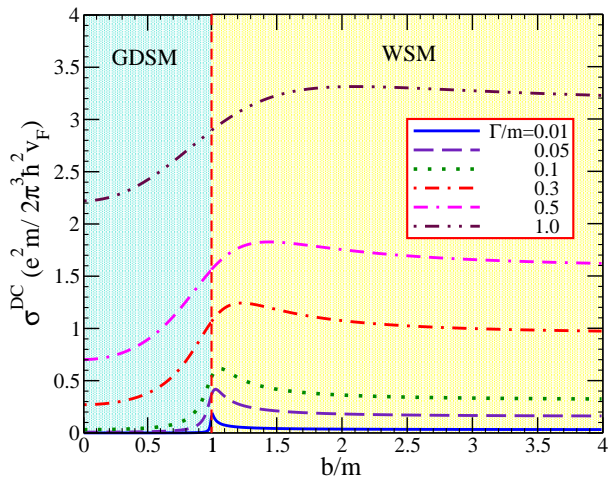} 
\vspace{0.5cm}
\caption{(Color online) The D.C. conductivity $\sigma^{DC}$ at charge neutrality in units of $\frac{e^2m}{2\pi^3\hbar^2v_{F}}$ as a function of the Zeeman parameter 
$\bb=b/m$. Six values of quasiparticle scattering rate $\bGamma=\Gamma/m$ are shown, namely $\bGamma=0.01$(solid blue),$\bGamma=0.05$(dashed indigo), 
$\bGamma=0.1$(dotted green), $\bGamma=0.3$(dashed-dotted red), $\bGamma=0.5$(double dashed-dotted purple) and $\bGamma=1.0$(double dotted-dashed brown). The region of 
$b<m$ (light blue shading) corresponds to the gapped Dirac phase while for $b>m$ we have 
the Weyl phase (light green shading). The vertical dotted red line for $\bb=1$ is the phase boundary. The two insets are schematics of the carrier dispersion curve 
conduction band. In all cases the minimum D.C. conductivity has its maximum as a function of $\bb$ in the Weyl phase. The smaller the value of $\bGamma$ the closer this 
maximum is to the phase boundary and the larger is its relative value to the background at $\bb$ large. As $\bGamma$ increases the various curves fall above each other 
and the less pronounced is the drop in the gapped state. In our units the D.C. conductivity is given by the function $\bGamma^2\N(\bGamma,\bb)$ where $\N(\bGamma,\bb)$ 
is defined in Eq.~(\ref{mathfracN}).} 
\label{fig:Fig4}
\end{figure}

When both $m$ and $b$ are finite we need to return to Equs.~(\ref{DC-cond-final}) and (\ref{mathfracY-b0}). In Fig.~\ref{fig:Fig4} we plot the D.C. conductivity 
$\sigma^{DC}$ in units of($\frac{e^2 m}{2\pi^3\hbar^2v_F}$) as a function of $\bb=b/m$ for various values of the normalized quasiparticle scattering rate 
$\bGamma=\Gamma/m$. The boundary between Weyl semimetal $\bb>1$ (light yellow shaded region) and gapped Dirac regime $\bb<1$(light green shaded region) is indicated as a 
red dashed vertical line at $\bb=1$. The six values of $\bGamma$ shown are $\bGamma=0.01$(solid blue),$\bGamma=0.05$(dashed indigo), $\bGamma=0.1$(dotted green),
$\bGamma=0.3$(dashed-dotted red), $\bGamma=0.5$(double dashed-dotted purple) and $\bGamma=1.0$(double dotted-dashed brown). In our chosen units the function that is 
plotted is $\bGamma^2\N(\bGamma,\bb)$ where $\N(\bGamma,\bb)$ is given in Eq.~(\ref{mathfracN}). In the Weyl phase, well away from the phase boundary at $\bb=1$, all the 
curves shown become fairly constant i.e. the D.C. conductivity is pretty well independent of $b$ as we have anticipated. As the value of $\bb$ decreases, the D.C. 
conductivity shows an increase and this is more pronounced the smaller the value of $\bGamma$. Independent of the value of $\bGamma$, $\sigma^{DC}$ has a maximum before 
the phase boundary is reached and the closer it is to $\bb=1$ the smaller the value of $\bGamma$. As the boundary is crossed into the gapped Dirac phase the conductivity 
drops towards a small value as compared to its magnitude for $\bb\rightarrow \infty$ particularly when $\bGamma$ is itself small. As an example for $\bGamma=0.05$ it has 
a value of $\sim 0.045$ in units of $\frac{e^2 m}{2\pi^3\hbar^2v_F}$ as compared with 0.168 at $\bb=4$. The boundary between Weyl and gapped Dirac is better probed in 
transport in the limit when the scattering rate $\Gamma$ is small as compared with the characteristic gap scale $m$.

\begin{figure}[h]
\includegraphics[width=2.5in,height=3.2in, angle=270]{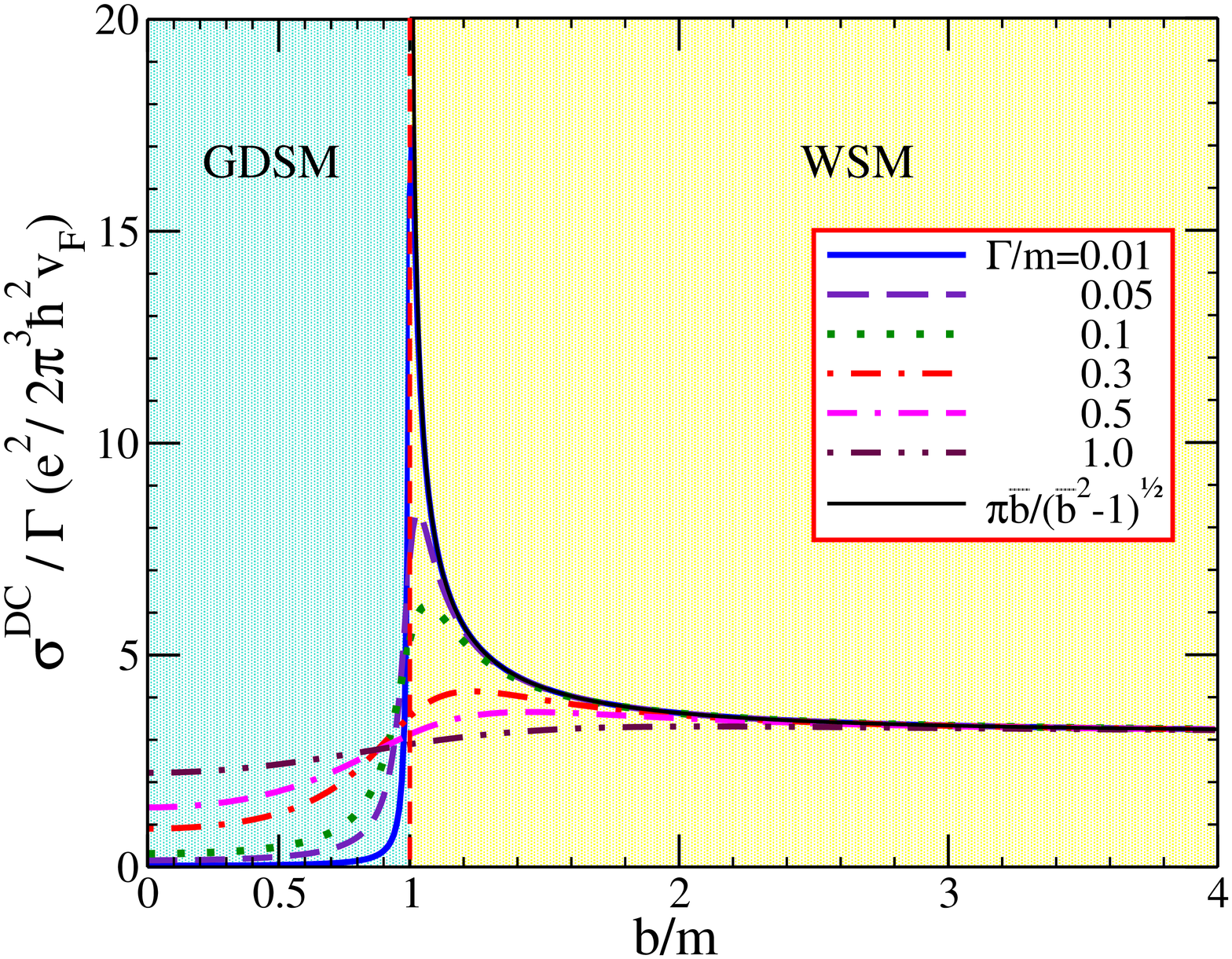} 
\vspace{0.5cm}
\caption{(Color online) The minimum D.C. conductivity normalized to $\Gamma$ units of $\frac{e^2}{2\pi^3\hbar^2v_{F}}$ as a function of the Zeeman parameter $\bb=b/m$ 
for the same values of quasiparticle scattering rate $\bGamma=\Gamma/m$ as shown in Fig.~\ref{fig:Fig4}. In the units chosen what is plotted is $\bGamma\N(\bGamma,\bb)$ 
of Eq.(\ref{gamma.mathfracN})with $\N(\bGamma,\bb)$ defined in Eq.~(\ref{mathfracN}). This function contains the factor of $\bGamma=\Gamma/m$ less than the function 
plotted in Fig.~\ref{fig:Fig4}. At the larger values of $\bb=b/m$ shown, all curves saturate to the same value of 3.4. This value is approximately $10\%$ larger than the 
$\bb=\infty$ limit which gives $\pi$. This function clearly shows that the value at the peaks in this family of curves increases sharply as $\bGamma=\Gamma/m$ is reduced 
and moves closer to the phase boundary. Also shown for comparison is the solid black curve which represent the function $\pi\bb/\sqrt{\bb^2-1}$.} 
\label{fig:Fig5}
\end{figure} 

\begin{figure}[h]
\includegraphics[width=2.5in,height=2.5in, angle=270]{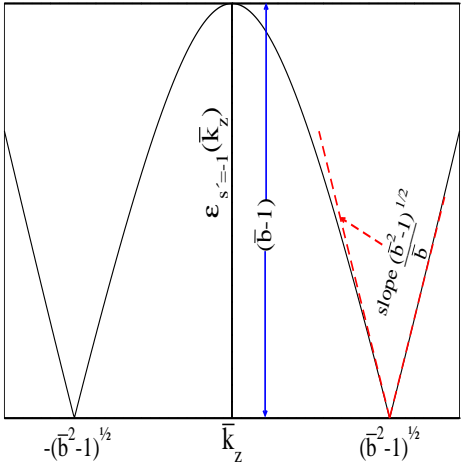} 
\vspace{0.5cm}
\caption{(Color online) Electronic dispersion $\bepsilon_{s'=-1}(\bk_{z})=|\sqrt{\bk^2_{z}+1}-\bb|=\epsilon_{1,-1}/m$ as a function of z-component of the momentum normalize 
by $m(\bk_{z}=k_{z}/m)$ for $k_{x}=k_{y}=0$. The height of the dome at $\bk_{z}=0$ is given by $\bb-1$ which is shown as the blue arrow and the Weyl nodes are situated 
at $\bk_{z}=\pm\sqrt{\bb^2-1}$ (here $v_{F}=1$). The dashed red line shows the slope $\frac{\sqrt{\bb^2-1}}{\bb}$of $\bepsilon_{s'=-1}(\bk_{z})$ at the nodes. This slope 
decreases towards zero as we reduce $\bb$ towards 1 (phase boundary).} 
\label{fig:Fig6}
\end{figure}

In Fig.~\ref{fig:Fig5} we plot our results for the D.C. conductivity in a slightly different way. We return to Eq.~(\ref{DC-cond-final}) and note that the function
\be
\label{gamma.mathfracN}
\bGamma\N(\bGamma,\bb)\hspace{-0.1cm}=\hspace{-0.2cm}\int^{+\infty}_{0}\hspace{-0.5cm}dx \hspace{-0.1cm}\lb\frac{\bGamma}{\bGamma^2\hspace{-0.1cm}+\hspace{-0.1cm}\lp\hspace{-0.1cm}\sqrt{x^2\hspace{-0.1cm}+\hspace{-0.1cm}1}\hspace{-0.1cm}+\hspace{-0.1cm}\bb \rp^2}+
\frac{\bGamma}{\bGamma^2\hspace{-0.1cm}+\hspace{-0.1cm}\lp\hspace{-0.1cm}\sqrt{x^2\hspace{-0.1cm}+\hspace{-0.1cm}1}\hspace{-0.1cm}-\hspace{-0.1cm}\bb \rp^2}\rb
\ee
can be simplified in the clean limit.
For $\bGamma \to 0$ we can replace both Lorentzians in (\ref{gamma.mathfracN}) by Dirac delta functions and note that for $\bb>1$ in the Weyl phase the 
second integral will give zero while the first reduces to $\pi\int^{\infty}_0 dx \delta(\sqrt{x^2+1}-\bb)$ which gives $\pi\bb/\sqrt{\bb^2-1}$. This factor is related 
to the slope of the dispersion curves and is an effective Fermi velocity as we now illustrate. In Fig.~\ref{fig:Fig6} we show a schematic of the electronic dispersion 
curves. We consider $\k_x=\k_y=0$ and plot the dispersion curve associated with the $s'=-1$ branch only. This is the branch which features 
the Weyl nodes for $\bb>1$. In the figure we plot $\epsilon_{s'=-1}(k_z)/m$ which we denote by 
$\bepsilon_{s'=-1}$ as a function $\bk_z=k_z/m$. The height at $\bk_{z}=0$ of the dome is $\bb-1$ while there is a node at $\pm\sqrt{\bb^2-1}$ (solid black curve). Also 
shown as dashed red curves are the slopes out of the Weyl nodes which have the value $\sqrt{\bb^2-1}/\bb$. This is a critical factor in our model and shows that the effective 
Fermi velocity that is to be associated with the Weyl node is modified for the $k_z$ coordinate by a factor of $\sqrt{\bb^2-1}/\bb$ so that the effective Fermi velocity 
gets smaller as the GDSM boundary is approached from the WSM side. The factor $\sqrt{b^2-m^2}$ is well known from the anomalous Hall effect which is given \cite{Burkov} 
by the universal quantized value $\frac{e^2}{h}$ times the distance between the Weyl nodes. Here the related factor gives the important function $\bGamma\N(\bGamma,\bb)$ 
of Eq.~(\ref{gamma.mathfracN}). This function is plotted in Fig.~\ref{fig:Fig5}. What is plotted is the D.C. conductivity normalized to $\Gamma$ in units of 
$\frac{e^2}{2\pi^3\hbar^2v_F}$ as a function of $\bb$ for the same six values of $\bGamma$ which were used in Fig.~\ref{fig:Fig4}. We see that for 
$\bb \gtrsim 3$ all curves have merged and there is no dependence in this region on the scattering rate $\Gamma$. Further above $\bb \gtrsim 2$ the dependence on $\bb$ 
is very small and $b$ has essentially dropped out as we know it must for large $b$. The peak in the D.C. conductivity near the phase boundary is 
greatly enhanced as $\bGamma$ is reduced. In fact we have plotted in the same figure our analytic result for the limit $\bGamma\rightarrow 0$ namely 
$\pi\bb/\sqrt{\bb^2-1}$ as solid black curve which tracks well our numerical results. In this same limit the D.C. conductivity of Eq.(\ref{DC-cond-final}) reduce to 
$\sigma^{DC}=\frac{e^2 \Gamma}{2\pi^2\hbar^2v_{F}}\frac{\bb}{\sqrt{\bb^2-1}}$ which we recognize as being the same as for an isolated Dirac node of Ref.(\onlinecite{Nicol}) 
except that the Fermi velocity $v_{F}$ is to be replaced by its effective value $v_{F}\sqrt{\bb^2-1}/\bb$ (see Fig.~\ref{fig:Fig6}). This provides a simple explanation of 
why the minimum D.C. conductivity increases as the phase boundary between WSM and GDSM is approached. Finally returning to Eq.(\ref{gamma.mathfracN}) when $\bb<1$ (GDSM) 
and $\bGamma\to0$ neither Dirac delta function can contribute and we get $\sigma^{DC}_{min}=0$, as broadening is increased this region of course fills in 
(Fig.~\ref{fig:Fig5}).

\begin{figure}[h]
\includegraphics[width=2.5in,height=3.2in, angle=270]{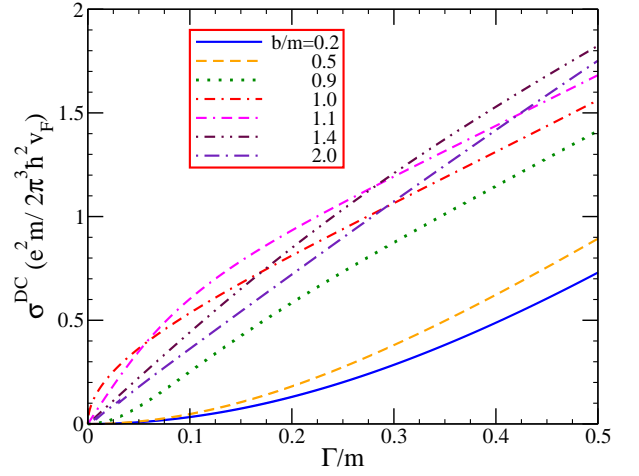} 
\vspace{0.5cm}
\caption{(Color online) The D.C. conductivity at charge neutrality in units of $\frac{me^2}{2\pi^3\hbar^2v_{F}}$ as a function of the quasiparticle scattering rate 
$\bGamma=\Gamma/m$ for seven values of the Zeeman parameter $\bb=b/m$, namely light blue ($\bb=0.2$), yellow($\bb=0.5$),green($\bb=0.9$), red($\bb=1.0$)(phase boundary),
purple($\bb=1.1$), brown($\bb=1.4$) and dark blue($\bb=2.0$). } 
\label{fig:Fig7}
\end{figure}

It is instructive to plot our results in yet another way. In Fig.~\ref{fig:Fig7} we present the D.C. conductivity in units of $\frac{e^2 m}{2\pi^3\hbar^2v_F}$ as a function 
of $\bGamma$ for seven values of the Zeeman parameter $\bb=b/m$, namely light blue ($\bb=0.2$), yellow($\bb=0.5$),green($\bb=0.9$), red($\bb=1.0$)(phase boundary), purple
($\bb=1.1$), brown($\bb=1.4$) and dark blue($\bb=2.0$). In the chosen units what is plotted is the function $\bGamma^2\N(\bGamma,\bb)$ with $\N(\bGamma,\bb)$ given in 
Eq.~(\ref{mathfracN}). In the Weyl semimetal phase the curves are linear in $\bGamma=\Gamma/m$ at small values of the scattering rate after which they show concave 
downward behavior. The range in $\bGamma=\Gamma/m$ over which the linearity hold increases with increasing value of $\bb=b/m$. In the gapped Dirac phase the behavior is 
concave upward. The red curve is at the phase boundary and has its own characteristic behavior.

\section{Lowest order finite temperature and doping correction to D.C. conductivity}
\label{sec:IV}

\begin{figure}[h]
\includegraphics[width=2.5in,height=3.2in, angle=270]{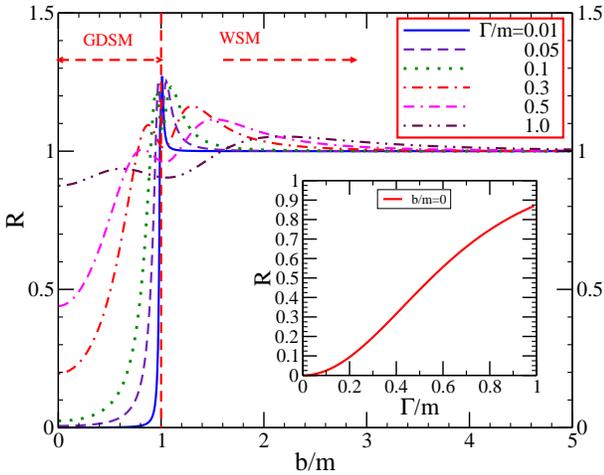} 
\vspace{0.5cm}
\caption{(Color online) The ratio $R=\bGamma^2\M(\bGamma,\bb)/\N(\bGamma,\bb)$ which defines the coefficient of the correction term to the D.C. conductivity for finite 
$\mu$ or $T$ in Eq.(\ref{27b}) shown as a function of $b/m$ for 6 values of the constant residual scattering rate $\Gamma/b$ as labeled in the figure. The inset gives 
$R$  as a function of $\Gamma/m$ for the case $b/m=0$.}
\label{fig:Fig8}
\end{figure}

Next we work out the lowest order correction to the D.C. conductivity for finite temperature and chemical potential $\mu$ away from the charge neutrality point, assuming 
$T/\Gamma$ and $\mu/\Gamma$ to be much less than one. These quantities define the approach to the minimum conductivity of the previous section. To accomplish this we 
return to Eq.(\ref{OCD}) and (\ref{OCIB}), take the limit $\Omega\rightarrow 0$ and introduce polar coordinates from ($k_x, k_y$) variables to get,
\bea
\label{sigma-finite-T-and-doping}
&& \hspace{-0.3cm}\sigma^{tot}(T,\mu)= \frac{e^2}{\hbar^2 v_{F}}\hspace{-0.1cm}\int^{+\infty}_{0} \hspace{-0.3cm}\frac{\rho d\rho}{2\pi^3}\hspace{-0.1cm} \int^{+\infty}_{0} \hspace{-0.5cm}dk_z  
\hspace{-0.2cm}\sum_{s'=\pm}\hspace{-0.1cm} \int^{+\infty}_{-\infty}\hspace{-0.5cm} d\omega\hspace{-0.1cm}\lp \hspace{-0.1cm}-\frac{\partial f(\omega)}{\partial \omega}\hspace{-0.1cm}\rp\hspace{-0.1cm}\times \nonumber\\
&& \Gamma^2 \biggl[(1-\frac{\rho^2}{2\epsilon^2_{s'}}) \frac{2}{\lp\Gamma^2+(\omega-\epsilon_{s'})^2\rp\lp\Gamma^2+(\omega+\epsilon_{s'})^2\rp} +\nonumber\\ 
&& \hspace{-0.3cm}\frac{\rho^2}{2\epsilon^2_{s'}}\biggl\{\lp\frac{1}{\Gamma^2+(\omega+\epsilon_{s'})^2}\rp^2 \hspace{-0.2cm}+ \hspace{-0.1cm}\lp\frac{1}{\Gamma^2+(\omega-\epsilon_{s'})^2}\rp^2 \biggr\} \biggr]
\eea
with $\epsilon^2_{s'}=\rho^2+\lp\sqrt{k^2_z+m^2}+bs'\rp^2$. In the zero temperature limit $-\frac{\partial f(\omega)}{\partial \omega}$ is a Dirac delta function 
$\delta(\omega)$ for $\mu=0$ and Eq.~(\ref{sigma-finite-T-and-doping}) reduces to Eq.~(\ref{DC-conductivity}) of the previous section. We are interested in the 
lowest order correction for finite $T$ and/or $\mu$ when these energies are small as compared with the quasiparticle scattering rate $\Gamma$. For this purpose it is 
sufficient to expand the Lorentzians in Eq.~(\ref{sigma-finite-T-and-doping}) to the order $\omega^2$. After considerable but straightforward algebra we obtain,
\bea
\label{sigma-tot}
&& \hspace{-0.3cm}\sigma^{tot}(T,\mu)= \frac{e^2 \Gamma^2}{\pi^3\hbar^2 v_{F}}\hspace{-0.1cm}\sum_{s'=\pm}\hspace{-0.1cm}\int^{+\infty}_{0}\hspace{-0.5cm}\rho d\rho \hspace{-0.1cm} \int^{+\infty}_{0} \hspace{-0.5cm}dk_z  
\hspace{-0.2cm}\int^{+\infty}_{-\infty}\hspace{-0.5cm} d\omega\hspace{-0.1cm}\lp \hspace{-0.1cm}-\frac{\partial f(\omega)}{\partial \omega}\hspace{-0.1cm}\rp\hspace{-0.1cm}\times \nonumber\\
&& \lb A_{s'}(\rho,k_z,\Gamma) + B_{s'}(\rho,k_z,\Gamma) \omega^2\rb
\eea
with
\be
\label{As'}
A_{s'}(\rho,k_z,\Gamma)=\lp \frac{1}{\Gamma^2+\epsilon^2_{s'}}\rp^2
\ee
and 
\be
\label{Bs'}
B_{s'}(\rho,k_z,\Gamma)=2\lp \frac{1}{\Gamma^2+\epsilon^2_{s'}}\rp^4 \lb\epsilon^2_{s'}-\Gamma^2+2\rho^2\rb
\ee
The integral over $\omega$ in Eq.~(\ref{sigma-tot}) can be done and gives
\bea
\label{sigma-tot-new}
&& \hspace{-0.3cm}\sigma^{tot}(T,\mu)= \frac{e^2 \Gamma^2}{\pi^3\hbar^2 v_{F}}\hspace{-0.1cm}\int^{+\infty}_{0}\hspace{-0.5cm}\rho d\rho \hspace{-0.1cm} \int^{+\infty}_{0} \hspace{-0.5cm}dk_z  
\times \nonumber\\
&& \lb A(\rho,k_z,\Gamma) + B(\rho,k_z,\Gamma)\lp \mu^2 + \frac{\pi^2}{3}T^2 \rp \rb
\eea
with $A(\rho,k_z,\Gamma)=\sum_{s'=\pm}A_{s'}(\rho,k_z,\Gamma)$ and $B(\rho,k_z,\Gamma)=\sum_{s'=\pm}B_{s'}(\rho,k_z,\Gamma)$. It is convenient to introduce 
$\alpha^2_{s'}=\lp\sqrt{k^2_z+m^2} + bs'\rp^2$ and to note that $\epsilon^2_{s'}=\rho^2+\alpha^2_{s'}$. The integration over $\rho$ can be done 
analytically and as $\alpha^2_{s'}$ is independent of $\rho$, it is to be treated as a constant. All required integrals have the form,
\be
\label{Int-general-form}
\int^{+\infty}_{0} \hspace{-0.5cm}\frac{\rho d\rho}{\lp\Gamma^2+\rho^2+\alpha^2_{s'}\rp^{n}} =\frac{1}{2\lp\Gamma^2+\alpha^2_{s'}\rp^{n-1}} \lp\frac{1}{n-1}\rp
\ee
Defining,
\be
\label{A}
A=\hspace{-0.1cm}\int^{+\infty}_{0}\hspace{-0.5cm}\rho d\rho \hspace{-0.1cm} \int^{+\infty}_{0} \hspace{-0.5cm}dk_z  A(\rho,k_z,\Gamma)=\lp\frac{1}{2m}\rp\N(\bGamma,\bb)
\ee
where $\bGamma=\Gamma/m, \bb=b/m$ and $\N(\bGamma,\bb)$ is the function defined in the previous section (Eq.~(\ref{mathfracN})) and we recover Eq.~(\ref{DC-cond-final}) 
for the D.C. conductivity at zero temperature as we must. Here we are interested in the correction term for finite $T$ and $\mu$. We define,
\be
\label{B}
B=\hspace{-0.1cm}\int^{+\infty}_{0}\hspace{-0.5cm}\rho d\rho \hspace{-0.1cm} \int^{+\infty}_{0} \hspace{-0.5cm}dk_z  B(\rho,k_z,\Gamma)=\lp\frac{1}{6m^3}\rp\M(\bGamma,\bb)
\ee
with the new function 
\bea
\label{mathfracM}
&& \M(\bGamma,\bb)=\int^{+\infty}_{0} dx \biggl[ \frac{5}{\biggl\{\bGamma^2+\lp\sqrt{x^2+1}+\bb\rp^2\biggr\}^2}-  \nonumber \\
&& \hspace{-0.3cm} \frac{4\bGamma^2}{\biggl\{\bGamma^2+\lp\sqrt{x^2+1}+\bb\rp^2\biggr\}^3} + \frac{5}{\biggl\{\bGamma^2+\lp\sqrt{x^2+1}-\bb\rp^2\biggr\}^2}- \nonumber \\
&& \hspace{-0.3cm} \frac{4\bGamma^2}{\biggl\{\bGamma^2+\lp\sqrt{x^2+1}-\bb\rp^2\biggr\}^3}\biggr]
\eea
and its contribution to the D.C. conductivity $\sigma^{DC}(\bGamma,\bb)$ is 
\be
\frac{e^2\Gamma^4}{6\pi^3\hbar^2v_{F}} \M(\bGamma,\bb) \lp \frac{\mu^2}{\Gamma^2}+\frac{\pi^2}{3}\frac{T^2}{\Gamma^2} \rp \nonumber
\ee
and
\bea
&& \sigma^{tot}=\frac{e^2 m}{2\pi^3\hbar^2v_{F}} \biggl[\bGamma^2 \N(\bGamma,\bb)+\frac{1}{3} \bGamma^4\M(\bGamma,\bb)\biggl\{\lp\frac{\mu}{\Gamma}\rp^2+ \nonumber\\
&& \frac{\pi^2}{3}\lp\frac{T}{\Gamma}\rp^2\biggr\}\biggr]
\label{27b}
\eea
The ratio $R=\bGamma^2\M(\bGamma,\bb)/\N(\bGamma,\bb)$ is equal to 1 for independent ungapped Dirac nodes. We already saw in previous section that when $m=0$ in our model, 
the quantity $m\bGamma^2\N(\bGamma,\bb)$ becomes independent of $b$ and is equal to $\pi\Gamma$. The quantity $m\bGamma^4\M(\bGamma,\bb)$ can be written for $m\rightarrow 0$
in the form,
\bea
\label{30}
&&\hspace{-0.5cm} m\bGamma^4\M(\bGamma,\bb)=\int^{+\infty}_{0} dx \biggl[ \frac{5\Gamma^4}{\biggl\{\Gamma^2+\lp x + b\rp^2\biggr\}^2}-  \nonumber \\
&& \hspace{-0.5cm} \frac{4\Gamma^6}{\biggl\{\Gamma^2+\lp x+b\rp^2\biggr\}^3} + \frac{5\Gamma^4}{\biggl\{\Gamma^2+\lp x- b\rp^2\biggr\}^2}- \nonumber \\
&& \hspace{-0.5cm} \frac{4\Gamma^6}{\biggl\{\Gamma^2+\lp x-b\rp^2\biggr\}^3}\biggr]
\eea
We can change variable $x+b$ to $y$ in the first two terms and $x-b$ to $y$ in the second pair to get
\bea
&& \int^{+\infty}_{-b} dy \lb \frac{5\Gamma^4}{\lp\Gamma^2+ y^2\rp^2}-\frac{4\Gamma^6}{\lp \Gamma^2+y^2\rp^3}\rb + \nonumber \\
&& \int^{+\infty}_{b} dy \lb \frac{5\Gamma^4}{\lp\Gamma^2+ y^2\rp^2}-\frac{4\Gamma^6}{\lp \Gamma^2+y^2\rp^3}\rb
\eea
Both integrals can be done analytically, the contribution from the lower limits cancel out and we are left with $\pi\Gamma$ independent of $b$. Thus for $m=0$ the ratio of 
interest $R=1$. For finite $m$ but $b=0$ we again get an analytic result. For $m\bGamma^2\N(\bGamma,b=0)$ we had $\frac{\pi \Gamma^2}{\sqrt{\Gamma^2+m^2}}$. For 
$m\bGamma^4\M(\bGamma,b=0)$ we get
\be
m\bGamma^4\M(\bGamma,b=0)= \frac{\pi}{2} \frac{\Gamma^4\lp 5m^2+2\Gamma^2\rp}{\lp m^2+\Gamma^2\rp^{5/2}}
\ee
So that in the gapped Dirac case the ratio $R$ is not one but rather is given by
\be
\label{R-Dirac}
R=\frac{\Gamma^2\lp 5m^2 +2 \Gamma^2 \rp}{2\lp m^2+\Gamma^2\rp^2}
\ee
which goes like $\frac{5}{2} \lp\frac{\Gamma^2}{m^2}\rp$ for $\frac{\Gamma}{m}<<1$. The ratio $R$ is plotted in Fig.~\ref{fig:Fig5} as a function of $b/m=\bb$ for 
several values of $\bGamma=\Gamma/m$. 
Solid blue curve for $\bGamma=0.01$, dashed orange is for $\bGamma=0.05$, dotted green for $\bGamma=0.1$, dashed-dotted red for 
$\bGamma=0.3$, double dashed-dotted purple for $\bGamma=0.5$ and dashed-double dotted brown is for $\bGamma=1.0$.
In the Weyl semimetal phase $R$ is never far from one. In the clean limit $\bGamma\rightarrow 0$ we can easily see from Eq.~(\ref{30}) that $\bGamma^4\M(\bGamma,\bb)$ 
will behave like $\frac{\pi \bb}{\sqrt{\bb^2-1}}$ which is the same as for $\bGamma^2\N(\bGamma,\bb)$ and hence we get exactly one. 
In this same limit , $R=0$ in the GDSM phase and $R$ is close to the solid black curve of Fig.~\ref{fig:Fig8}. As $\Gamma$ increases out of the clean limit the magnitude 
of $R$ gradually increases in the GDSM phase, the peaks at the phase boundary is reduced and moves further to the right (i.e. to larger values of $\bb$ in the WSM phase). 
At $\bGamma=1.0$ (dashed-double dotted brown curve) deviations from $R=1$ are now small for any value of $\bb$.
At large $\bb>>1$  we recover as we expect the gapless Dirac point node result which is again one. In the gapped Dirac state $R$ can deviate strongly from one as we have shown 
analytically. The case $\bb=0$ is shown in the inset to Fig.~\ref{fig:Fig8} where we see that it rises quadratically as a function of $\Gamma/m$ out of zero and has 
reached $7/8$ by $\bGamma=1$ as we expect from Eq.(\ref{R-Dirac}).

\section{Finite photon energy approach to D.C. conductivity at charge neutrality}
\label{sec:V}

In the Appendix we provide a formula for the finite frequency optical conductivity at zero temperature after the integration over the intermediate frequency $\omega$ of 
Eq.~(\ref{OCD}), for the intraband piece, and Eq.~(\ref{OCIB}) for the interband optical transitions has been done analytically. If we are only interested in the lowest order contribution 
for finite photon energy $\Omega$, we can use the expansions given in Eq.(\ref{mathfracY-expansion}) and (\ref{mathfracH-expansion}) to get the lowest correction to 
$\sigma_{xx}(T=0,\Omega)$ which is quadratic in $\Omega$ to get 
\bea
\label{delta-sigmaD}
&& \delta \sigma^{D}(T=0,\Omega)= \frac{e^2 \Gamma}{2\pi^3\hbar^2 v_{F}} \sum_{s'=\pm} \int^{+\infty}_{0}\hspace{-0.1cm} d\tk_z \int^{+\infty}_{0} \hspace{-0.1cm}\trho d\trho \times \nonumber \\
&& \lp\frac{\trho^2}{2\tepsilon^2_{s'}}\rp \frac{4\lp 2\tepsilon^2_{s'}-1\rp}{3\lp \tepsilon^2_{s'}+1 \rp^4} \tOmega^2
\eea
and 
\bea
\label{delta-sigmaIB}
&& \delta \sigma^{IB}(T=0,\Omega)= \frac{e^2 \Gamma}{2\pi^3\hbar^2 v_{F}} \sum_{s'=\pm} \int^{+\infty}_{0}\hspace{-0.1cm}d\tk_z \int^{+\infty}_{0}\hspace{-0.1cm}\trho d\trho \times \nonumber \\
&& \lp1-\frac{\trho^2}{2\tepsilon^2_{s'}}\rp \frac{4\lp 4\tepsilon^2_{s'}-1\rp}{3\lp \tepsilon^2_{s'}+1 \rp^4} \tOmega^2
\eea
where all twiddle variables mean we have divided by the quasiparticle scattering rate. The total contribution to the conductivity is,
\bea
\label{delta-sigma-tot}
&&\delta\sigma^{tot}(T=0,\Omega)=\delta \sigma^{D}+\delta \sigma^{IB}\nonumber \\
&& =\frac{2e^2 \Gamma\tOmega^2}{3\pi^3\hbar^2 v_{F}} \sum_{s'=\pm} \int^{+\infty}_{0}\hspace{-0.1cm}d\tk_z \int^{+\infty}_{0}\hspace{-0.1cm}\trho d\trho 
\biggl\{\frac{3}{\lp \tepsilon^2_{s'}+1 \rp^3}-\nonumber \\
&& \frac{\lp4-\tilde{\alpha}^2_{s'}\rp}{\lp \tepsilon^2_{s'}+1 \rp^4}\bigg\}
\eea
where $\tilde{\alpha}^2_{s'}=\lp\sqrt{\tk^2_z+\tm^2} + s'\tb\rp^2$ and $\tepsilon^2_{s'}=\trho^2+\tilde{\alpha}^2_{s'}$. 

\begin{figure}[h]
\includegraphics[width=2.5in,height=3.2in, angle=270]{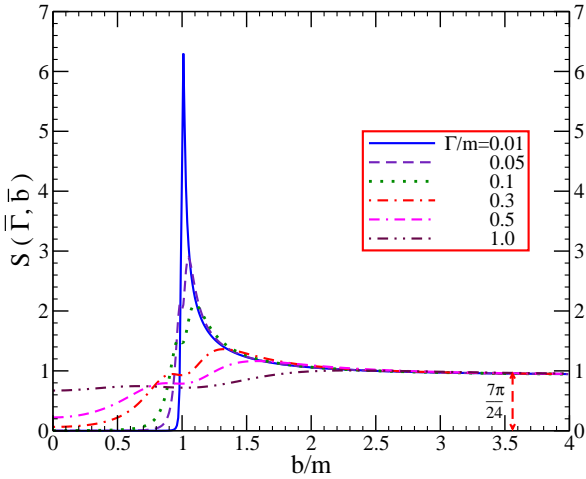} 
\vspace{0.5cm}
\caption{(Color online) The function $S(\bGamma,\bb)$ defined in Eq.(\ref{mathfrakG-new}) shown for 6 values of $\Gamma/m$, $\Gamma/m=0.01$ solid blue curve, dashed 
orange for $\Gamma/m=0.05$, dotted green for $\Gamma/m=0.1$, dashed-dotted red for $\Gamma/m=0.3$, dotted-double dashed purple for $\Gamma/m=0.5$, dashed-double dotted 
brown for $\Gamma/m=1.0$. The function $S(\bGamma,\bb)$ gives the coefficient for the approach to the minimum D.C. conductivity as photon energy $\Omega\to 0$. The lead term is quadratic in 
$\Omega/\Gamma$ (here $\mu=T=0$).} 
\label{fig:Fig9}
\end{figure}

As before the integration over $\trho$ can be done analytically to get 
\bea
\label{delta-sigma-tot-new}
&& \frac{\delta\sigma^{tot}(T=0,\Omega)}{\Gamma}=\frac{e^2 \tOmega^2}{3\pi^3\hbar^2 v_{F}} \hspace{-0.2cm}\sum_{s'=\pm} \int^{+\infty}_{0}\hspace{-0.5cm}d\tk_z 
\biggl[ \frac{11}{6\lp 1+\tilde{\alpha}^2_{s'}\rp^2}-\nonumber \\
&& \frac{5}{3\lp 1+\tilde{\alpha}^2_{s'}\rp^3} \biggr ]
\eea
which can be rewritten in terms of the band variables i.e. $\bar{a}=a/m$ for any variable $a$ to get 
$\frac{\delta\sigma^{tot}(T=0,\Omega)}{\Gamma}=\frac{e^2 \tOmega^2 \bGamma^3}{3\pi^3\hbar^2 v_{F}} \G(\bGamma,\bb)$ with 
\bea
\label{mathfrakG-new}
&& S(\bGamma,\bb)\equiv\bGamma^3\G(\bGamma,\bb)= \nonumber \\
&& \sum_{s'=\pm} \int^{+\infty}_{0}dx \biggl[\frac{11\bGamma^3}{6\lp\bGamma^2+\lp\sqrt{x^2+1}+s'\bb\rp^2\rp^2}-\nonumber \\
&& \frac{5\bGamma^5}{3\lp\bGamma^2+\lp\sqrt{x^2+1}+s'\bb \rp^2\rp^3}\biggr]
\eea
Note that in the case $m=b=0$, $\delta\sigma^{tot}(T=0,\Omega)$ must reduce to the known result for a doubly degenerate Dirac point node. Our Eq.~(\ref{mathfrakG-new})
reduces in this limit to,
\be
\frac{e^2 \Gamma \tOmega^2}{3\pi^3\hbar^2 v_{F}} \int^{+\infty}_{0}\hspace{-0.5cm}d k_z 
\biggl[\frac{11\Gamma^3}{6\lp\Gamma^2+k^2_{z}\rp^2}-\frac{5\Gamma^5}{3\lp\Gamma^2+k^2_{z}\rp^3}\biggr]
\ee
which simplifies to
\be
\label{38}
\frac{7}{72}\lp\frac{e^2}{\pi^2\hbar^2 v_{F}}\rp\tOmega^2
\ee
In Ref.[\onlinecite{Nicol}] the conductivity of a single Dirac node for $m=0$ is given as the sum of the equations (\ref{delta-sigmaD}) and (\ref{delta-sigmaIB}) and 
to the first correction for finite $\Omega$ equals our quoted result Eq.(\ref{38}) after accounting for the two nodes. This serves as a check on our calculations. The 
function $\bGamma^3\G(\bGamma,\bb)$ defined by Eq.~(\ref{mathfrakG-new}) is plotted in Fig.~\ref{fig:Fig9} as a function of $\bb$ for various values of $\bGamma$. For 
large values of $\bb$ it saturate to a value of $\frac{7\pi}{24}$ independent of the value of the residual scattering rate $\bGamma$ employed, as is also the case for 
the results presented in Fig.~\ref{fig:Fig5} for the related function $\bGamma\N(\bGamma,\bb)$ of Eq.(\ref{gamma.mathfracN}). Just as in that case, as $\bb$ decreases 
towards the GDSM-WSM phase boundary at $\bb=b/m=1$, the function $S(\bGamma,\bb)$ of Fig.~\ref{fig:Fig9} increases with the magnitude at the maximum larger as $\bGamma$ 
decreases and its position along the $b/m$ axis decreasing and becoming one in the limit $\bGamma\to 0$. Finally as the phase boundary to the gapped state is crossed 
$S(\bGamma,\bb)$ is rapidly reduced towards zero for small $\bGamma$ while for $\bGamma=\Gamma/m=1$ it remains of order one up to $\bb=0$.

\section{Conclusion}
\label{sec:VI}

Within a continuum model Hamiltonian in which a doubly degenerate Dirac node is split into two Weyl points through broken time reversal symmetry we calculated 
A.C. optical response and D.C. transport within a Kubo formalism. The model Hamiltonian has a gap term ($m$) and a Zeeman term ($b$) and exhibit two distinct phases. A 
Weyl phase for $b/m>1$ and a gapped Dirac phase (GDSM) for $b/m<1$. In the Weyl phase (WSM) only one of the two branches of the electronic dispersion curves is ungapped 
with the second branch is gapped at $m+b$. In the GDSM both branches have gaps of ($m-b$) and ($m+b$) respectively. For the ungapped Weyl branch the slope of the 
dispersion curves
out of the node is $v_{F} \sqrt{b^2-m^2}/b$ and decreases as $b/m$ decreases towards 1. There is also a dome of height ($b-m$) along the $k_{z}$ axis for $k_{x}=k_{y}=0$.
and associated Van Hove singularity. In the clean limit \cite{Tabert} this translates into a kinks at $\Omega=2(b-m)$ in $\sigma_{xx} (T=0, \Omega)$. There is a sharp 
sincrease in A.C. conductivity at 
$\Omega=2(b+m)$ from the onset of the second gapped branch. The conductivity shows quasilinear behavior in the range $\Omega<2(b-m)$ and a second  quasilinear region with 
smaller slope in the range $2(b-m)<\Omega<2(b+m)$. Here we find that  the introduction of residual scattering ($\Gamma$) broadens out these structures. However for 
$\Gamma/m=0.05$ the two distinct quasilinear regions remain and the imprint of finite $b$ and $m$ remains in $\sigma_{xx}(T=0, \Omega)$. For the $m=b=0$ the optical response 
is, in contrast, linear except very near to $\Omega=0$ where it bends over to intercept the vertical axis at some small but finite value of the D.C. conductivity. The result 
$\sigma^{DC}=\frac{e^2 \Gamma}{2\pi^2\hbar^2v_{F}}\times \frac{\bb}{\sqrt{\bb^2-1}}$ is the same as for two independent Dirac nodes \cite{Nicol} except that it is the 
effective Fermi velocity out of the Weyl node in our model Hamiltonian ($v_{F}\frac{\sqrt{\bb^2-1}}{\bb}$) which replaces $v_{F}$. This identifies the mechanism for the 
increase in $\sigma^{DC}$ as $\bb\to 1$. When $\Gamma$  is increased to $\Gamma/m=0.4$ however there remains little sign of two separate quasilinear regions and the 
conductivity has evolved towards its value in the $m=b=0$ case with some minor quantitative but no qualitative differences.

When finite values of doping (chemical potential $\mu$) are considered the main modifications to the conductivity arise in the range $\Omega<2\mu$. The interband transitions 
are Pauli blocked and the optical spectral weight is transferred to the intraband transitions which manifest as a Drude peak. The region of reduced conductivity between 
Drude and remaining interband background provides information on the magnitude of the chemical potential $\mu$ as well as the magnitude of the scattering rate $\Gamma$. 
If $\Gamma<m$ and $2\mu$ is less than $2(b-m)$ clear signatures of finite $m$ and $b$ can remain in $\sigma_{xx}(T=0, \Omega)$ above the Pauli blocked region.

In the WSM phase the leading correction to the D.C. conductivity due to finite temperature $T$, dopping $\mu$ and photon frequency $\Omega$ are found to be quadratic in 
$T/\Gamma$, $\mu/\Gamma$ and $\Omega/\Gamma$ respectively. The coefficient of these quadratic laws normalized to the value of the D.C. limit are one for $b/m>1$, and 
increase as $b/m$ is reduced towards the WSM-GDSM boundary. These deviations from one are never large and $\lesssim 20\%$. In the GDSM phase they rapidly drop towards zero 
for $\Gamma/m<1$ which is characteristic of a gapped state. For $b=0$ we find the normalized coefficient for the $T/\Gamma$ and/or $\mu/\Gamma$ dependence equal to 
$\frac{\Gamma^2\lp 5m^2 +2 \Gamma^2 \rp}{2\lp m^2+\Gamma^2\rp^2} $ which is 1 as $\Gamma>m$. The D.C. conductivity itself is modified by a factor 
$\Gamma/\sqrt{\Gamma^2+m^2}$ with no change in Fermi velocity which remains $v_{F}$.

\subsection*{Acknowledgments}
Work supported in part by the Natural Sciences and Engineering Research Council of Canada (NSERC) and by the Canadian Institute for Advanced Research (CIFAR).

\appendix

\section{}

The dynamic optical conductivity $\sigma_{xx}(T,\Omega)$ at temperature $T$ and photon energy $\Omega$ has two contributions, the intraband or Drude contribution 
$\sigma^{D}_{xx}(T,\Omega)$ is given \cite{Tabert} by
\bea
\label{OCD}
&& \sigma^{D}_{xx}(T, \Omega)= \frac{e^2\pi}{\hbar^2 v_{F}} \sum_{ss'=\pm}\int^{+\infty}_{-\infty}\hspace{-0.6cm} d\omega \lb \frac{f(\omega)-f(\omega+\Omega)}{\Omega}\rb \times \nonumber \\ 
&&  \int\hspace{-0.2cm} \frac{d^3\k}{(2\pi)^3} \lp \frac{k^2_{\perp}}{2\epsilon^2_{ss'}(\k)}\rp A(\epsilon_{ss'},\omega)A_{ss'}(\epsilon_{ss'},\omega+\Omega),
\eea
where $k^2_{\perp}=k^2_{x}+k^2_{y}$, $f(\omega)$ is the Fermi Dirac distribution function,
\be
\label{Fermi-function}
f(\omega)=\frac{1}{e^{\frac{\omega-\mu}{T}}+1}
\ee
with $\mu$ the chemical potential and $A(\epsilon_{ss'},\omega)$ is the carrier spectral density which for a constant quasiparticle scattering rate $\Gamma$ is a 
Lorentzian
\be
\label{spectral-density}
A(\epsilon_{ss'},\omega)= \frac{1}{\pi} \frac{\Gamma}{\Gamma^2+\lp \omega - \epsilon_{ss'}\rp^2}
\ee
In Ref.[\onlinecite{Tabert}] only the clean limit was considered in which case $A(\epsilon_{ss'},\omega)$ is a simple Dirac delta function $\delta(\epsilon_{ss'}-\omega)$ 
(see their Eq.~(A12)). The second contribution to the conductivity $\sigma^{IB}_{xx}(T,\Omega)$ comes from the interband optical transitions and has the form 
(Eq.~(A13) in Ref.[\onlinecite{Tabert}])
\bea
\label{OCIB}
&& \sigma^{IB}_{xx}(T, \Omega)= \frac{e^2\pi}{\hbar^2 v_{F}} \sum_{ss'=\pm}\int^{+\infty}_{-\infty}\hspace{-0.6cm} d\omega \lb \frac{f(\omega)-f(\omega+\Omega)}{\Omega}\rb \times \nonumber \\ 
&& \hspace{-0.5cm} \int\hspace{-0.2cm} \frac{d^3\k}{(2\pi)^3} \lp 1-\frac{k^2_{\perp}}{2\epsilon^2_{ss'}(\k)}\rp A(-\epsilon_{ss'},\omega)A_{ss'}(\epsilon_{ss'},\omega+\Omega).
\eea
At zero temperature the thermal factors in (\ref{OCD}) and (\ref{OCIB}) are Heaviside theta function which restrict the integration over $\omega$ to the interval 
($\mu-\Omega,\mu$). As noted in Ref.[\onlinecite{Mukherjee}] the integration over $\omega$ involves only the product of the spectral functions which can be done analytically 
and the Drude and interband conductivity written in terms of two explicit functions. For the interband part,
\be
\label{IntIB}
\frac{\sigma^{\hspace{-0.1cm}IB}_{\hspace{-0.1cm}xx}\hspace{-0.1cm} \lp\hspace{-0.1cm}T\hspace{-0.1cm}=\hspace{-0.1cm}0,\hspace{-0.07cm}\Omega\rp}{\Gamma}\hspace{-0.1cm}=\hspace{-0.1cm} 
\frac{e^2}{2\pi^3 \hbar^2v_{F}}\hspace{-0.15cm}\sum_{s'} \hspace{-0.2cm}\int^{\hspace{-0.05cm}\infty}_{\hspace{-0.05cm}0}\hspace{-0.2cm}
\frac{d\tk_{z}}{\tOmega}\hspace{-0.2cm}\int^{\hspace{-0.05cm}\infty}_{\hspace{-0.05cm}0}\hspace{-0.35cm} \trho d\trho 
\hspace{-0.1cm}\lp\hspace{-0.15cm}1\hspace{-0.1cm}-\hspace{-0.1cm}\frac{\brho^2}{2\tepsilon^2_{s'}}\hspace{-0.15cm}\rp \hspace{-0.13cm}\Y(\tmu,\tOmega,\tb,\tm,\tepsilon_{s'}), 
\ee
where we have introduced polar coordinates for $k_{x},k_{y}$ variables and have divided all variables by the scattering rate which has had the effect of scaling out 
$\Gamma$. Of course it remains in $\widetilde{\Omega}\equiv \frac{\Omega}{\Gamma}, \widetilde{\mu}\equiv \frac{\mu}{\Gamma}$,$\widetilde{m}\equiv \frac{m}{\Gamma}$ and 
$\widetilde{b}\equiv \frac{b}{\Gamma}$ while the other variables are dummies of integration. The function $\Y(\tmu,\tOmega,\tb,\tepsilon_{s'})$ has the form 
\bea
\label{Special-function-interband}
&& \Y(\tmu,\tOmega,\tb,\tm,\tepsilon_{s'})= \frac{1}{(\tOmega+2\tepsilon_{s'}) \lb 4 + \lp \tOmega + 2\tepsilon_{s'} \rp ^2 \rb} \times \nonumber  \\ 
&& \biggl [\ln \lp \frac{1+(\tmu+\tOmega+\tepsilon_{s'})^2}{1+(\tmu-\tepsilon_{s'})^2} \times \frac{1+(\tmu-\tOmega-\tepsilon_{s'})^2}{1+(\tmu+\tepsilon_{s'})^2}\rp 
+\nonumber  \\
&& (\tOmega+2\tepsilon_{s'}) \{ \arctan(\tmu+\tOmega+\tepsilon_{s'}) + \arctan(\tmu-\tepsilon_{s'}) - \nonumber  \\
&& \arctan(\tmu+\tepsilon_{s'}) - \arctan(\tmu-\tOmega-\tepsilon_{s'}) \} \biggr] +\nonumber  \\
&& \frac{1}{(\tOmega-2\tepsilon_{s'}) \lb 4 - \lp \tOmega + 2\tepsilon_{s'} \rp ^2 \rb} \times \nonumber  \\ 
&& \biggl [\ln \lp \frac{1+(\tmu+\tOmega-\tepsilon_{s'})^2}{1+(\tmu+\tepsilon_{s'})^2} \times \frac{1+(\tmu-\tOmega+\tepsilon_{s'})^2}{1+(\tmu-\tepsilon_{s'})^2}\rp 
+\nonumber  \\
&& (\tOmega-2\tepsilon_{s'}) \{ \arctan(\tmu+\tOmega-\tepsilon_{s'}) + \arctan(\tmu+\tepsilon_{s'}) - \nonumber  \\
&& \arctan(\tmu-\tepsilon_{s'}) - \arctan(\tmu-\tOmega+\tepsilon_{s'}) \} \biggr].
\eea
For the intraband case we obtain,
\be
\label{OC Drude}
\frac{\sigma^{D}_{xx}(T\hspace{-0.1cm}=\hspace{-0.1cm}0,\Omega)}{\Gamma}=\frac{e^2}{2\pi^3 \hbar^2v_{F}}\hspace{-0.15cm} \sum_{s'}\hspace{-0.15cm} 
\int^{\infty}_{0} \hspace{-0.15cm}\frac{d\tk_{z}}{\tOmega} \hspace{-0.2cm}\int^{\infty}_{0}\hspace{-0.4cm} \trho d\trho 
\frac{\trho^2}{2\tepsilon^2_{s'}} \mathfrak{H}(\tmu,\tOmega,\tb,\tm,\tepsilon_{s'}), 
\ee
with 
\bea
\label{Special-function-intraband}
&& \mathfrak{H}(\tmu,\tOmega,\tb,\tm,\tepsilon_{s'}) = \frac{1}{\tOmega \lb \tOmega^2 + 4\rb} \biggl [ \ln \biggl( \frac{\lp\hspace{-0.1cm}1\hspace{-0.1cm}+\hspace{-0.1cm}(\tmu\hspace{-0.05cm}+\hspace{-0.05cm}\tOmega\hspace{-0.05cm}+\hspace{-0.05cm}\tepsilon_{s'})^2\hspace{-0.1cm}\rp}{1\hspace{-0.1cm}+\hspace{-0.1cm}(\tmu\hspace{-0.05cm}+\hspace{-0.05cm}\tepsilon_{s'})^2} \times \nonumber \\ 
&& \hspace{-0.3cm}\frac{\lp\hspace{-0.1cm}1\hspace{-0.1cm}+\hspace{-0.1cm}(\tmu\hspace{-0.05cm}+\hspace{-0.05cm}\tOmega\hspace{-0.05cm}-\hspace{-0.05cm}\tepsilon_{s'})^2\hspace{-0.1cm}\rp}{1\hspace{-0.1cm}+\hspace{-0.1cm}(\tmu\hspace{-0.05cm}+\hspace{-0.05cm}\tepsilon_{s'})^2} 
\frac{\lp\hspace{-0.1cm}1\hspace{-0.1cm}+\hspace{-0.1cm}(\tmu\hspace{-0.05cm}-\hspace{-0.05cm}\tOmega\hspace{-0.05cm}+\hspace{-0.05cm}\tepsilon_{s'})^2\hspace{-0.1cm}\rp}{1\hspace{-0.1cm}+\hspace{-0.1cm}(\tmu\hspace{-0.05cm}-\hspace{-0.05cm}\tepsilon_{s'})^2}
\frac{\lp\hspace{-0.1cm}1\hspace{-0.1cm}+\hspace{-0.1cm}(\tmu\hspace{-0.05cm}-\hspace{-0.05cm}\tOmega\hspace{-0.05cm}-\hspace{-0.05cm}\tepsilon_{s'})^2\hspace{-0.1cm}\rp}{1\hspace{-0.1cm}+\hspace{-0.1cm}(\tmu\hspace{-0.05cm}-\hspace{-0.05cm}\tepsilon_{s'})^2}\hspace{-0.05cm} \biggr)\hspace{-0.05cm}+\hspace{-0.05cm} \nonumber \\
&&\tOmega \{\arctan(\tmu\hspace{-0.05cm}+\hspace{-0.05cm}\tOmega\hspace{-0.05cm}+\hspace{-0.05cm}\tepsilon_{s'})\hspace{-0.1cm}+\hspace{-0.1cm}\arctan(\tmu\hspace{-0.05cm}+\hspace{-0.05cm}\tOmega\hspace{-0.05cm}-\hspace{-0.05cm}\tepsilon_{s'})\hspace{-0.05cm}- \hspace{-0.05cm}\nonumber \\
&& \arctan(\tmu\hspace{-0.05cm}-\hspace{-0.05cm}\tOmega\hspace{-0.05cm}+\hspace{-0.05cm}\tepsilon_{s'})\hspace{-0.1cm}-\hspace{-0.1cm}\arctan(\tmu\hspace{-0.05cm}-\hspace{-0.05cm}\tOmega\hspace{-0.05cm}-\hspace{-0.05cm}\tepsilon_{s'}) \} \biggr].
\eea
Here $\tepsilon_{s'}=\sqrt{\trho^2 + \lp\sqrt{\tk^2_{z}+\tm^2}+s'\tb\rp^2}$. We will also be interested in the small $\tOmega$ limit of these functions. To order $\tOmega^4$ 
we have,
\bea
\label{mathfracY-expansion}
&& \frac{1}{\tOmega} \Y(\tmu,\tOmega,\tb,\tm,\tepsilon_{s})= \frac{2}{\tOmega(1+\tepsilon_{s}^2)^2}+\frac{4(4\tepsilon_{s}^2-1)\tOmega}{3(1+\tepsilon_{s}^2)^4}+\nonumber\\
&& \frac{2(93\tepsilon_{s}^4-150\tepsilon_{s}^2+13)\tOmega^3}{30(1+\tepsilon_{s}^2)^6} +O[\tOmega]^5
\eea
 
\bea
\label{mathfracH-expansion}
&& \frac{1}{\tOmega} \mathfrak{H}(\tmu,\tOmega,\tb,\tm,\tepsilon_{s})=\frac{2}{\tOmega(1+\tepsilon_{s}^2)^2}+\frac{4(2\tepsilon_{s}^2-1)\tOmega}{3(1+\tepsilon_{s}^2)^4}+\nonumber\\
&& \frac{(45\tepsilon_{s}^4-102\tepsilon_{s}^2+13)\tOmega^3}{15(1+\tepsilon_{s}^2)^6} +O[\tOmega]^5 
\eea

\end{document}